\documentclass[aps,prx,a4paper,twocolumn,superscriptaddress,showkeys,longbibliography]{revtex4-1}

\usepackage[colorlinks=true,bookmarks=false,citecolor=blue,linkcolor=red,hyperfootnotes=true,urlcolor=blue]{hyperref}

\usepackage{physics}
\usepackage{graphicx}
\usepackage{dcolumn}
\usepackage{bm}
\usepackage[dvipsnames]{xcolor}
\usepackage{wasysym}
\usepackage{lipsum}
\usepackage{enumitem}
\usepackage{booktabs} 
\usepackage{gensymb}
\usepackage{xcolor}
\usepackage{amssymb}
\usepackage{booktabs}
\usepackage{array}
\newcolumntype{L}{>{$}l<{$}}
\newcolumntype{C}{>{$}c<{$}} 

\newcolumntype{R}{>{$}r<{$}}

\usepackage{mathtools}
\usepackage{siunitx}
\usepackage{soul}
\usepackage{float}

\makeatletter
\def\p@subsection{}
\makeatother

\newcommand{\bfit}[1]{\boldsymbol{#1}}

\DeclareMathOperator{\sgn}{sgn}

\begin{document}

\title{Bipolaronic high-temperature superconductivity}

\author{C. Zhang}\thanks{These authors contributed equally: C. Zhang, J. Sous.} 
\affiliation{State Key Laboratory of Precision Spectroscopy,
East China Normal University, Shanghai 200062, China}

\author{J. Sous} \thanks{These authors contributed equally: J. Sous. C. Zhang}\thanks{Author to whom correspondence should be addressed; Present address: Department of Physics, Stanford University, Stanford, CA 93405, USA} \email{sous@stanford.edu}
\affiliation{Department of Physics, Columbia University, New York,
New York 10027, USA}

\author{D. R. Reichman} 
\affiliation{Department of Chemistry, Columbia University, New York,
New York 10027, USA}

\author{M. Berciu} 
\affiliation{Department of Physics and Astronomy, University of British Columbia, Vancouver, British Columbia V6T 1Z1, Canada}
\affiliation{Stewart Blusson Quantum Matter Institute, University of British Columbia, Vancouver, British Columbia V6T 1Z4, Canada}

\author{A. J. Millis} \thanks{Author to whom correspondence should be addressed} \email{ajm2010@columbia.edu}
\affiliation{Department of Physics, Columbia University, New York,
New York 10027, USA} 
\affiliation{Center for Computational Quantum Physics, Flatiron Institute, 162 5$^{th}$ Avenue, New York, NY 10010, USA}

\author{N. V. Prokof’ev}
\affiliation{Department of Physics, University of Massachusetts, Amherst, Massachusetts 01003, USA}

\author{B. V. Svistunov}
\affiliation{Department of Physics, University of Massachusetts, Amherst, Massachusetts 01003, USA}
\affiliation{Wilczek Quantum Center, School of Physics and Astronomy and T. D. Lee Institute, Shanghai Jiao Tong University, Shanghai 200240, China}

\date{\today}

\begin{abstract}
Electron-lattice interactions play a prominent role in quantum materials, making a deeper understanding of  direct routes to phonon-mediated high-transition-temperature ($T_{\mathrm{c}}$)  superconductivity desirable. However, it has been known for decades that weak electron-phonon coupling gives rise to low values of $T_{\mathrm{c}}$, while strong electron-phonon coupling leads to lattice instability or formation of bipolarons, generally assumed to be detrimental to superconductivity. Thus, the route to high-$T_{\mathrm{c}}$ materials from phonon-mediated mechanisms has heretofore appeared to be limited to raising the phonon frequency as in the hydrogen sulfides. Here we present a simple model for phonon-mediated high-$T_{\mathrm{c}}$ superconductivity  based on superfluidity of {\em light} bipolarons. In contrast to the widely studied Holstein model where lattice distortions modulate the electron's potential energy, we investigate the situation where lattice distortions modulate the electron hopping. This physics gives rise to small-size, yet light bipolarons, which we study using an exact sign-problem-free quantum Monte Carlo approach, demonstrating a new route to phonon-mediated high-$T_\mathrm{c}$ superconductivity.   We find that  $T_\mathrm{c}$ in our model generically and significantly exceeds typical upper bounds based on Migdal-Eliashberg theory or superfluidity of Holstein bipolarons. The key ingredient in this bipolaronic mechanism that gives rise to high $T_\mathrm{c}$ is the combination of light mass and  small size of bipolarons. Our work establishes principles  towards the design of high-$T_{\mathrm{c}}$ superconductors via  functional material engineering.
\end{abstract}

\maketitle

\section{Introduction}

The quest for new pathways to high-temperature superconductivity from physically simple ideas has captivated researchers for decades.  Conventional superconductivity is well understood within the standard framework of  the Bardeen-Cooper-Schrieffer (BCS) theory, in which the exchange of phonons between electrons acts as a pairing glue to produce superconductivity with a transition-temperature $T_\mathrm{c}$ that is a small fraction of the phonon frequency $\Omega$ ($\hbar=1$). In this framework, high $T_\mathrm{c}$ can arise in systems with very large phonon frequencies, and indeed the pressure-stabilized hydrides with very large phonon frequencies exhibit superconductivity at remarkably high temperatures~\cite{HighPressure1,HighPressure2}. A great deal of experimental and theoretical work has also focused on routes to high $T_\mathrm{c}$ based on unconventional superconductivity with non-$s$-wave pairing symmetry arising from electronic correlations~\cite{HighTcReview}.

The pursuit of phonon-mediated high-temperature superconductivity  from strong electron-phonon coupling at fixed phonon frequency $\Omega$ is, however, challenged by several important constraints. In a conventional superconducting material, such as Al, phonon-induced attraction between electrons induces a Cooper instability, giving rise to superconductivity with a $T_{\mathrm{c}}/\Omega$ that is small and vanishes as   the strength of the electron-phonon coupling $\lambda\rightarrow 0$. As $\lambda$ increases, $T_\mathrm{c}/\Omega$ increases.  However, the standard theoretical treatment of the large-$\lambda$ problem is based on the Migdal-Eliashberg approximation which breaks down for $\lambda$ larger than a critical value of order $1$~\cite{ChakravertyBipolaron,ME_CDW1,ME_CDW2,CohenBounds,MEbreakdownAlexandrov,MEBreakdownMillis,MEBreakdownKivelson} because of lattice reconstruction or formation of heavy bipolarons~\cite{ChakravertyBipolaron,MEbreakdownAlexandrov,CohenBounds,MEBreakdownKivelson}. In high-electron-density materials, increasing $\lambda$ beyond $\lambda \approx 1$ typically leads to lattice reconstruction into a new structure with a reduced $\lambda$, leading to a maximum value of $T_\mathrm{c}$ at $\lambda\approx 1$. At lower carrier density, the electron-lattice interaction may not induce lattice reconstruction; instead, bipolarons emerge and these bipolarons either form a charge-localized non-superconducting state~\cite{ChakravertyBipolaron,ME_CDW1,ME_CDW2} or undergo a superfluid transition at a $T_\mathrm{c}$ determined by the inverse of their effective mass. However, the effective mass of strongly-bound bipolarons has generically been believed to be large and to increase rapidly with $\lambda$~\cite{CRFBipolaron,BoncaHBipolaron,MacridinBipolaron}, implying  generically low values of $T_{\mathrm{c}}$ from bipolaronic superconductivity~\cite{CRFBipolaron}.

\begin{figure}[!t]
\raggedright
\includegraphics[width=0.92\columnwidth]{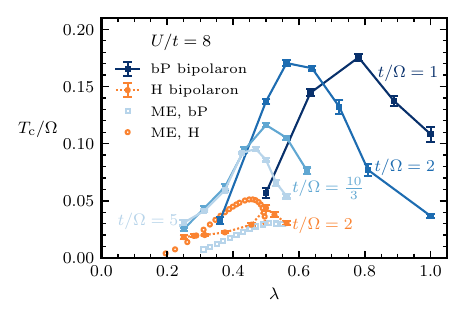}
\vspace{-4mm}
\caption{{\bf Bipolaronic high-$T_{\mathrm{c}}$ superconductivity.} $T_\mathrm{c}$ of the bond-Peierls (bP) bipolaronic superconductor (filled squares, solid blue lines) in units of the phonon frequency $\Omega$ for different adiabaticity ratios of the electron hopping $t$ to $\Omega$  with an onsite Hubbard repulsion $U = 8t$ as a function of the electron-phonon coupling $\lambda$ computed according to Eq.~\eqref{Eq:Tc} from QMC simulations of the bipolaron effective mass $m^{\! \star}_\mathrm{BP} \coloneqq [({\partial^2 E_{\mathrm{BP}} (K)}/{\partial K^2})|_{K=0}]^{-1}$ and its mean squared-radius $R^2_{\mathrm{BP}}\coloneqq \bra{\Psi_{\mathrm{BP}}} \hat{R}^2 \ket{\Psi_{\mathrm{BP}}}$. Here $E_{\mathrm{BP}}(K)$ is the bipolaron dispersion, $\Psi_{\mathrm{BP}}$ is the bipolaron ground state wavefunction, and $K$ is the bipolaron momentum. We contrast the behavior of the bipolaronic superconductivity in the bond-Peierls model against superconductivity of Holstein (H) bipolarons (filled circles, dotted orange line) for $t/\Omega = 2$ and $U=8t$ and against the prediction of Migdal-Eliashberg (ME) theory of strong-coupling superconductivity in the bond-Peierls (bP; empty squares) and Holstein (H; empty circles) models.  Here we use $\lambda = \alpha^2/2\Omega t$ for bond-Peierls bipolarons, $\lambda = \alpha^2_\mathrm{H}/8\Omega t$ for Holstein bipolarons, $\lambda =  8\alpha^2/2\pi\Omega t$ in Migdal-Eliashberg theory of the bond-Peierls model, and  $\lambda=\alpha^2_\mathrm{H}/2\pi\Omega t$ in Migdal-Eliashberg theory of the Holstein model, where $\alpha$ and $\alpha_\mathrm{H}$ are the electron-phonon coupling constants of the bond-Peierls and Holstein models, respectively, see Appendix~\ref{Appendix:AppA} for more details about the conventions used. Error bars represent statistical errors in QMC simulations corresponding to one standard deviation. This  comparison illustrates that $T_{\mathrm{c}}$ of the bond-Peierls bipolaronic superconductor can exceed previously expected upper bounds for phonon-mediated superconductivity in a wide swath of parameter space.}
\vspace{-4mm}
\label{fig:Fig1}
\end{figure}

Here, we challenge the widely held view that bipolaron formation is not favorable for high transition-temperature superconductivity by providing a concrete, experimentally relevant model for phonon-mediated bipolaronic high-$T_{\mathrm{c}}$ superconductivity with a $T_{\mathrm{c}}$ that can be significantly higher than previously  established upper bounds, see Fig.~\ref{fig:Fig1}. Our work is based on the observation that in the dilute limit bipolarons are in effect interacting bosons, with a transition temperature that depends both on the mass and the density. At fixed mass, the transition temperature increases as the density increases---until either the transition temperature becomes of the order of the bipolaron binding energy or the density becomes large enough that  the bipolarons significantly overlap, at which point the theory breaks down and, we suspect, the superconducting transition temperature saturates. Thus the maximum transition temperature is set by a combination of binding strength,  inverse mass and  inverse size, with small-size, light-mass, strongly bound bipolarons optimizing the maximum transition temperature.  In the extreme strong-coupling regime, the size saturates to a value of the order of the lattice constant, while it appears that in all models the  polaronic mass enhancement grows exponentially in $\lambda$. While these qualitative considerations are generic, the specifics and hence the maximum value of $T_\mathrm{c}$ will depend on the specifics of the underlying microscopic model studied. General understanding of this physics has been obtained from studies of the Holstein model, in which lattice distortions couple to the electron density (potential energy), and the focus has been on the  bipolaron mass, with the size receiving less attention. In these models the mass increases very rapidly as $\lambda$ is increased. Fr{\"o}hlich or extended-Holstein models in which the coupling to the electron density is longer ranged have also been studied; light masses have been found in circumstances involving an interplay of competing forces~\cite{PachaPolaron,BoncaEH,Pacha1}, but recent studies indicate that this light mass occurs in a very limited region of parameter space  to be relevant in realistic systems~\cite{Chandler}. However, alternative forms of electron-phonon coupling are also important.  In particular, any material with a unit cell consisting of more than a single atom will experience distortions of its atomic bonds that  locally modulate the electronic hopping, as in the Peierls~\cite{BarisicPeierls1} or Su-Schrieffer-Heeger (SSH) models~\cite{su1979solitons}. Models of this type exhibit significant differences in polaron formation relative to Holstein-type models~\cite{Capone1,Capone2,marchand2010sharp,ChaoPeierlsPolaron,CarbonePeierlsPolaron,SousBipolaron} including, importantly, lighter polarons~\cite{marchand2010sharp,ChaoPeierlsPolaron,CarbonePeierlsPolaron} and, as shown in Ref.~\cite{SousBipolaron}, lighter bipolarons. This previous analysis of bipolarons~\cite{SousBipolaron}, however, did not address superconductivity explicitly, did not examine the bipolaron size systematically and was limited to the one-dimensional case and constrained to the regime $t \sim \Omega$ ($t$ is the amplitude of the electronic hopping) leaving the possibility that mass enhancement would become more severe in higher dimensions in the adiabatic limit $t \gg \Omega$ relevant to most materials, thus destroying any hope of phonon-induced high-$T_\mathrm{c}$ behavior.

\begin{figure}[!t]
\raggedright
\includegraphics[width=0.92\columnwidth]{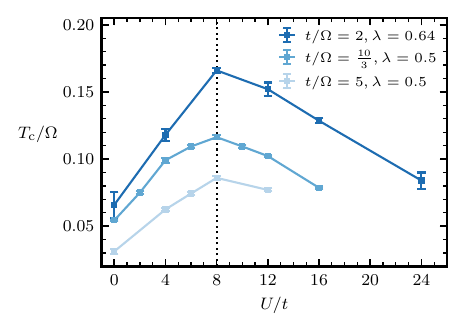}
\vspace{-4mm}
\caption{{\bf Coulomb repulsion-mediated enhancement of bipolaronic high-$T_{\mathrm{c}}$ superconductivity.} $T_\mathrm{c}$ of the bond-Peierls bipolaronic superconductor in units of the phonon frequency $\Omega$ for different adiabaticity ratios of the electron hopping $t$ to $\Omega$ at intermediate coupling $\lambda \sim 0.5 \; \mbox{-} \; 0.6$ as a function of the  onsite Hubbard repulsion $U$ in units of $t$ computed according to Eq.~\eqref{Eq:Tc} from QMC simulations of the bipolaron effective mass $m^{\! \star}_\mathrm{BP}$ and mean squared-radius $R^2_{\mathrm{BP}}$. Error bars represent statistical errors in QMC simulations corresponding to one standard deviation.  $T_\mathrm{c}$ of the bond-Peierls bipolaronic superconductor exceeds the largest value of $T_\mathrm{c}\sim 0.05 \Omega$ predicted by strong-coupling Migdal-Eliashberg theory (see Fig.~\ref{fig:Fig1}) even for large values of $U/t$.}
\label{fig:Fig2}
\vspace{-4mm}     
\end{figure}

In this work, we use a recently developed  numerically exact, sign-problem-free quantum Monte Carlo (QMC) approach~\cite{QMCBondBipolaron} to study bipolaronic superconductivity in a minimal model of Peierls electron-lattice coupling in two dimensions (2D). We compute  bipolaron properties in various regimes of coupling and adiabaticity (including deep in the adiabatic limit) and combine these results with analytical understanding of the bipolaron superfluid  phase diagram to determine $T_{\mathrm{c}}$ at which a liquid of  bipolarons undergoes a Berezinskii-Kosterlitz-Thouless  transition into a superfluid.  Our main results are as follows.  First, any non-zero $\lambda$ mediates an attractive interaction between electrons, giving rise to an $s$-wave bipolaronic superconductor with a $T_\mathrm{c} / \Omega$ that can become significantly larger than the upper bound predicted from Migdal-Eliashberg theory of strong-coupling superconductivity or from Holstein bipolaron superconductivity nearly everywhere in parameter space, see Fig.~\ref{fig:Fig1}.   Second, we find that Coulomb repulsion, modeled phenomenologically as a Hubbard $U$ term, \emph{enhances} the magnitude of  $T_{\mathrm{c}}$ of the $s$-wave bipolaronic superconductor up to a  critical value of $U/t$, beyond which $T_{\mathrm{c}}$ becomes suppressed (at intermediate to strong coupling, this behavior extends to  large values of $U/t$), see Fig.~\ref{fig:Fig2}. Finally, in our theory $T_{\mathrm{c}}$ is largest when $t/\Omega \sim 1\; \mbox{-}\;2$, implying that manipulating the  stiffness of a crystal via structural or moir\'e engineering or fabricating crystals with light atoms offers a path towards realizing high-temperature superconductors. Some aspects of the physics we discuss here may be operative in known materials, possibly including the iron-based pnictide superconductors.

\section{Formalism}

\subsection{Model of bond-phonon-coupled electrons}

We consider a minimal model of Peierls electron-phonon coupling, the bond-Peierls model (also referred to as the bond-SSH model) on a 2D square lattice. In this model the electronic hopping between two sites is modulated by an oscillator associated with the bond connecting the two sites. This coupling can arise if  transverse oscillations of out-of-plane atoms modulate the hopping of electrons between atoms in the plane. This is believed to occur in the superconducting iron pnictide materials, as discussed in Section~\ref{Outlook} and Appendix~\ref{Appendix:AppG}. The   Hamiltonian is
\begin{equation}
\hat{\cal H}=\hat{\cal H}_\mathrm{e} + \hat{\cal H}_\mathrm{ph} + \hat{\cal{V}}_\mathrm{{e\mbox{-}ph}}.
\label{Eq:H}
\end{equation}
Here electrons with spin $\sigma \in \{\uparrow,\downarrow\}$ are governed by a Hubbard model $\hat{\cal H}_\mathrm{ e} =-t \sum_{\langle i,j \rangle, \sigma}^{}\left( \hat{c}_{i,\sigma}^\dagger \hat{c}_{j,\sigma} + \mathrm{h.c.}\right) + U \sum_{i}^{}\hat{n}_{i,\uparrow} \hat{n}_{i,\downarrow}$
with onsite repulsion $U$ and  $\hat n_{i,\sigma} = \hat{c}_{i,\sigma}^\dagger \hat{c}_{i,\sigma}$ at site $i$. The notation $\langle i,j \rangle$ refers to nearest-neighbor sites. We set the lattice constant $a=1$ in what follows. We model  distortions of the bonds connecting sites $i$ and $j$ as Einstein oscillators $\hat{\cal H}_\mathrm{ph} = \sum_{\langle i,j \rangle} \big(\frac{1}{2} K \hat{X}_{i,j}^2 + \hat{P}_{i,j}^2/2M \big) = \Omega \sum_{\langle i,j \rangle} \hat{b}_{i,j}^\dagger \hat{b}_{i,j}$ with frequency  ($\hbar=1$) $\Omega = \sqrt{K/M}$ (note $\hat{X}_{i,j}$ is a single oscillator associated with the bond, not a composite variable representing a difference of displacements of the atoms at the two ends of the bond). We take the interaction between electrons and phonons
\begin{eqnarray}
\hat{\cal{V}}_\mathrm{{e\mbox{-}ph}} &=& \tilde{\alpha}  \sum_{\langle i,j \rangle,\sigma}^{}\left( \hat{c}_{i,\sigma}^\dagger \hat{c}_{j,\sigma} +
\mathrm{h.c.}\right) \hat{X}_{i,j} \nonumber \\
&=& \alpha\sum_{\langle i,j \rangle,\sigma}^{}\left( \hat{c}_{i,\sigma}^\dagger \hat{c}_{j,\sigma} +
\mathrm{h.c.}\right)\left( \hat{b}_{i,j}^\dagger+ \hat{b}_{i,j}\right)
\label{Eq:Veph}
\end{eqnarray}
to be the simplest coupling term within the family of Peierls models describing the modulation of electron hopping by an oscillator $\hat{X}_{i,j}$ associated with the bond connecting sites $i$ and $j$ with coupling coefficient $\alpha = \tilde{\alpha}/{\sqrt{2M\Omega}}$. We henceforth set $M=1$. The relevant parameters are a dimensionless coupling constant $\lambda = (\tilde{\alpha}^2/K)/4t = \alpha^2/(2\Omega t)$, the ratio of the typical polaronic energy scale to the free electron energy scale, and an adiabaticity parameter $t/\Omega$.  As discussed in Appendix~\ref{Appendix:AppB}, this model does not have a sign problem in the singlet two-electron sector (unlike other, related models, e.g. the site-Peierls model~\cite{SousBipolaron}) and therefore can be studied to great accuracy using quantum Monte Carlo methods.

\begin{figure*}
\centering
\includegraphics[width=1.778\columnwidth]{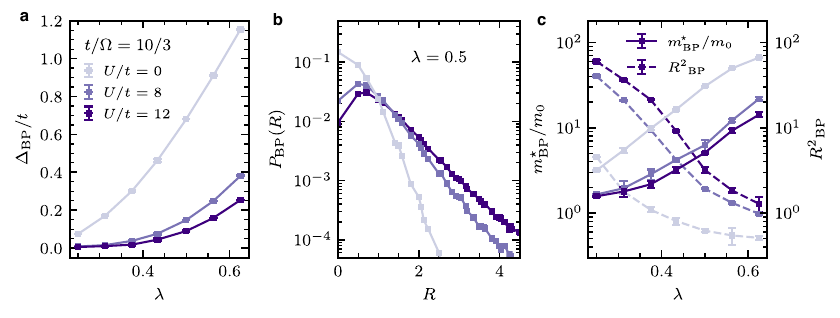}
\vspace{-4mm}
\caption{{\bf Bipolaron properties in the bond-Peierls model.} Bipolaron properties computed from QMC calculations performed on Eq.~\eqref{Eq:H} at adiabaticity parameter $t/\Omega = 10/3$ as a function of the electron-phonon coupling  $\lambda = \alpha^2/(2\Omega t)$ for different onsite Hubbard $U$ (in units of the electron hopping $t$): {\bf a}. Bipolaron binding energy $\Delta_{\mathrm{BP}}$ in units of the electron hopping $t$, {\bf b}. Bipolaron radial size probability density distribution (absolute value squared of the bipolaron wavefunction) $P_\mathrm{BP}(R)$ for $\lambda =0.5$, and {\bf c}. Bipolaron effective mass $m^{\! \star}_\mathrm{BP}$ in units of the mass of two free electrons $m_0 = 2m_e = 1/t$ and its mean squared-radius $R^2_{\mathrm{BP}}$. Error bars represent statistical errors in QMC simulations corresponding to one standard deviation. Error bars in $P_\mathrm{BP}(R)$ correspond to statistical errors smaller than the symbol size and therefore are not shown.}\label{fig:Fig3}
\vspace{-4mm}
\end{figure*}

\subsection{Method}

Using a QMC approach based on a path-integral formulation of the electronic sector combined with either a real-space diagrammatic or a Fock-space path-integral representation of the phononic sector~\cite{QMCBondBipolaron} allows us to study pairing and singlet bipolaron formation in the two-electron sector of the model. The absence of a sign problem, specific to this particular microscopic formulation and sector of the model, enables numerically exact results with small statistical errors on large lattices even in the challenging regime of $t \gg \Omega$ all the way to the heavy mass limit, see Appendix~\ref{Appendix:AppB} for details. While the coupling of electronic hopping to phonons can take various forms, our approach here allows us to draw generic conclusions about high-$T_{\mathrm{c}}$ bipolaronic superconductivity due to the modulation of electronic hopping by lattice distortions in the entire parameter space.

\section{Superfluid of bipolarons}

In 2D, bipolarons undergo a superfluid transition at a temperature $T\leq T_\mathrm{c}$, where $T_{\mathrm{c}}$ is determined by the bipolaron density and effective mass and depends only double-logarithmically weakly on the effective bipolaron-bipolaron interactions~\cite{LogBose2D1,LogBose2D2,LogBose2D3}.   We can thus safely ignore bipolaron-bipolaron interactions, barring any competing instability, e.g. phase separation or Wigner crystallization. Based on prior work~\cite{NoceraPS} we argue that these instabilities are unlikely. These considerations reduce our problem to that of the superfluidity of a gas of hard-core bipolarons in 2D, for which
$T_\mathrm{c} \approx 1.84 \rho_\mathrm{BP}/m^{\! \star}_\mathrm{BP}$~\cite{LogBose2D3}, where $\rho_\mathrm{BP}$ is the density of bipolarons  and $m^{\!\star}_\mathrm{BP}$ is the bipolaron effective mass~\footnote{Since $T_\mathrm{c}$ depends double-logarithmically weakly on the effective bipolaron-bipolaron interaction, neglecting deviations from the hard-core bipolaron-bipolaron interaction in our estimates of $T_\mathrm{c}$ can only lead to small uncertainties~\cite{LogBose2D2,LogBose2D3}, which, nonetheless, do not affect any of our main results.}. This formula for $T_\mathrm{c}$ remains valid in a broad density range so long as bipolarons do not overlap. The largest  $T_\mathrm{c}$ from this mechanism thus arises for a $\rho_\mathrm{BP}$ corresponding to a liquid of bosons with an inter-particle separation that is on the order of the bipolaron radial size $R_\mathrm{BP}$, which, after lattice regularization,  must be at least unity, i.e. for $\rho_\mathrm{BP} = \min\{(1/(\pi R^2_\mathrm{BP}), 1/\pi)\}$. From this we obtain an estimate for the maximum $T_{\mathrm{c}}$ of the Berezinskii-Kosterlitz-Thouless transition of the bipolaronic liquid that depends only on the bipolaron properties given by
\begin{eqnarray}
T_\mathrm{c} \approx
    \begin{cases}
        \frac{0.5}{m^{\! \star}_\mathrm{BP} R^2_\mathrm{BP}} & \text{if $R^2_\mathrm{BP}\geq 1$}\\
        \frac{0.5}{m^{\! \star}_\mathrm{BP}} & \text{otherwise}\\
    \end{cases}
\label{Eq:Tc}    
\end{eqnarray}

\paragraph{Bipolaronic high-$T_\mathrm{c}$ superconductivity.} Figure~\ref{fig:Fig1}  presents $T_\mathrm{c}/\Omega$ computed from Eq.~\eqref{Eq:Tc} using $m^{\! \star}_\mathrm{BP}$ and $R^2_\mathrm{BP}$ obtained from QMC simulations (see Appendix~\ref{Appendix:AppB}\ref{Appendix:AppB2}) of Eq.~\eqref{Eq:H} as a function of $\lambda$ for different $t/\Omega$ at $U=8t$.  Our results shown in Fig.~\ref{fig:Fig1} prove that high-$T_{\mathrm{c}}$ bipolaronic superconductivity is not only possible, but is robust even in the presence of a large Coulomb repulsion $U=8t$. To appreciate this result, we contrast our computed $T_\mathrm{c} /\Omega$ against upper bounds based on superfluidity of Holstein bipolarons computed using the same methodology as for the bond-Peierls bipolarons or Migdal-Eliashberg theory of strong-coupling superconductivity out of a Fermi liquid applied to the bond-Peierls and Holstein models. We find that $T_\mathrm{c}$ of the bond-Peierls bipolaronic superconductor generically exceeds these bounds in a large swath of parameter space. $T_\mathrm{c}$ of the Holstein bipolaronic superconductor, for $t/\Omega = 2$ at $U=0$, rapidly drops with $\lambda$ from a maximum of $\sim 0.05\Omega$ at  $\lambda \sim 0.25$ (Appendix~\ref{Appendix:AppC2}). As $U/t$ increases, the bipolaron mass decreases~\cite{BoncaHBipolaron, macridin2003phonons}, but the binding energy drops very rapidly and the radius correspondingly increases~\cite{macridin2003phonons}, limiting $T_\mathrm{c}$ to even smaller values, see Appendix~\ref{Appendix:AppC2} for more details. In contrast, the bond-Peierls bipolaron becomes strongly bound but remains relatively light as $\lambda$ increases and can resist a large $U/t$, consistent with the predictions of Ref.~\cite{SousBipolaron}, see Fig.~\ref{fig:Fig3}. The comparison of our calculations to that of Migdal-Eliashberg theory (Appendix~\ref{Appendix:AppD}) reveals that $T_\mathrm{c}$ of the bond-Peierls bipolaronic superconductor also exceeds the maximum $T_\mathrm{c}$ corresponding to strong-coupling superconductivity out of a Fermi liquid. This maximum $T_\mathrm{c}$  is in qualitative agreement with a typical upper bound  based on McMillan's  approach to conventional superconducting materials~\cite{McMillanFormula}.  McMillan's approach is valid only in the regimes of validity of Migdal-Eliashberg theory, i.e. up to $\lambda\approx 1$~\cite{CohenBounds,MEbreakdownAlexandrov,MEBreakdownKivelson,TcBoundKivelson}. Thus, a typical upper bound from the McMillan approach can be estimated at $\lambda = 1$ to give a maximum $T_\mathrm{c}/\Omega \sim 0.05$ for a Coulomb pseudopotential $\mu^\star = 0.12$ (see Appendix~\ref{Appendix:AppE}), in qualitative agreement with our results of  Migdal-Eliashberg theory applied to the two models presented in Fig.~\ref{fig:Fig1}. These comparisons illustrate that $T_{\mathrm{c}}$ obtained in our model of bipolaronic superconductivity significantly and generically exceed previously held expectations.

Figure~\ref{fig:Fig1} demonstrates a remarkable phenomenology of the high-$T_{\mathrm{c}}$ bipolaronic superconductivity. In particular, $T_\mathrm{c}/\Omega$ exhibits a non-monotonic, dome-like dependence on $\lambda$ with a peak that shifts to smaller values with larger $t/\Omega$.  We can understand this behavior as follows.  In the anti-adiabatic limit $t/\Omega \ll 1$, phonon exchange mediates an instantaneous pair-hopping interaction between electrons $\frac{-2 \alpha^2}{\Omega -U} \sum_{\langle i,j \rangle}\left( \hat{c}_{i,\uparrow}^\dagger \hat{c}_{i,\downarrow}^\dagger \hat{c}_{j,\downarrow} \hat{c}_{j,\uparrow}+ \mathrm{h.c.}\right)$, which induces formation of light-mass, strongly bound bipolarons with an $s$-wave wavefunction (see Appendix~\ref{Appendix:AppF} and Ref.~\cite{SousBipolaron}). By continuity, we expect this behavior to qualitatively persist and develop a frequency dependence (retardation) as $t/\Omega$ increases, accompanied by a proclivity for the bipolaron mass to increase as the number of phonons in the bipolaronic cloud grows. This competition between phonon-mediated kinetic energy-enhancing electron pair-hopping  interactions and a tendency to gain mass determines the properties of bipolaronic superconductivity. Our numerics reveal a parameterically large, physically relevant regime in which the bipolaron mass $m^{\! \star}_\mathrm{BP}$ exhibits weak to moderate enhancement whilst retaining a relatively small radial size $R^2_\mathrm{BP}$ and a large binding energy $\Delta_\mathrm{BP}$~\cite{SousBipolaron}, see Fig.~\ref{fig:Fig3}. This behavior is completely absent in the standard Holstein model in which bipolarons rapidly become heavy in a manner that depends exponentially on the electron-phonon coupling strength~\cite{BoncaHBipolaron,MacridinBipolaron}.  The characteristic behavior of our model, which we believe to hold generically for Peierls-coupled systems, explains the increase in $T_\mathrm{c}/\Omega$ with $\lambda$ up to an optimal $\lambda_{\mathrm{op}}$ beyond which bipolarons enter a regime of exponential mass enhancement that becomes prominent for $\lambda > \lambda_{\mathrm{op}}$ and larger $t/\Omega$. Nonetheless, over a broad swath of parameter space we find the simulated $T_\mathrm{c}/\Omega$ curves to surpass all previously held expectations.

\paragraph{Coulomb repulsion enhancement of $T_\mathrm{c}$ of bipolaronic superconductivity.} Most importantly, $T_\mathrm{c}/\Omega$ in Fig.~\ref{fig:Fig1} exceeds these bounds even for large values of $U/t$, demonstrating the robustness of the bipolaronic mechanism against Coulomb repulsion.  Figure~\ref{fig:Fig2} shows that $T_\mathrm{c}/\Omega$  exhibits a dome-like dependence also on $U/t$. The value of $U$ that maximizes $T_\mathrm{c}$ depends  on $\lambda$ (not shown).   At intermediate coupling $\lambda \sim 0.5 \; \mbox{-} \; 0.6$, $T_\mathrm{c}/\Omega$ peaks around a value of $U=8t$, which varies little with $t/\Omega$. This unconventional enhancement of the value of $T_{\mathrm{c}}$ of our $s$-wave bipolaronic superconductor up to such large values of $U/t$ can be understood from  Fig.~\ref{fig:Fig3}, which reveals that the bipolaron's effective mass $m^{\! \star}_\mathrm{BP}$ and its squared-radius $R^2_\mathrm{BP}$ depend on $U/t$, but in opposite ways. For larger $\lambda$, $R^2_\mathrm{BP}$ depends very weakly on $U/t$, while $m^{\! \star}_\mathrm{BP}$ decreases with increasing $U/t$, explaining the growth in $T_{\mathrm{c}}$.  This implies that in this limit the bipolaron size is already sufficiently large to enable the bound pair to avoid the Hubbard repulsion so that increasing $U$ bears no effect on the symmetry of the pairing wavefunction. However, the bipolaron binding energy $\Delta_\mathrm{BP}$ also decreases with increasing $U/t$, and this decrease ultimately limits $T_\mathrm{c}$, which cannot be greater than $\Delta_\mathrm{BP}$. In the Holstein model, the $U$ term directly competes with an onsite phonon-mediated attraction~\cite{BoncaHBipolaron,MacridinBipolaron} and while this results in a decrease in $m^{\! \star}_\mathrm{BP}$, it also causes a rapid increase in $R^2_\mathrm{BP}$ accompanied by a fast decrease in $\Delta_\mathrm{BP}$, and the latter two factors are more important and mean that ultimately $U$ does not significantly enhance $T_c$, see Appendix~\ref{Appendix:AppC2}. This analysis reveals that the Peierls bipolarons are generally much less sensitive to Coulomb repulsion than their Holstein counterparts~\cite{SousBipolaron}.

\begin{figure}[!t]
\raggedright
\includegraphics[width=0.92\columnwidth]{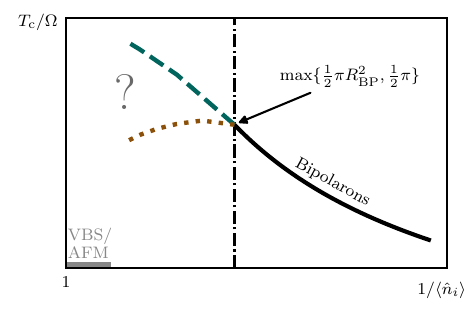}
\vspace{-4mm}
\caption{{\bf Fate of bipolaronic high-$T_\mathrm{c}$ superconductivity in the limit of large electronic densities.} Schematic diagram illustrating the expected dependence of the superconducting transition temperature $T_\mathrm{c}$ in the model of Eqs.~\eqref{Eq:H} - \eqref{Eq:Veph} on the average electronic density $\langle \hat{n}_i \rangle$. For $\langle \hat{n}_i \rangle \lesssim \min\{2/\pi R^2_\mathrm{BP},2/\pi\}$,  bipolarons form a dilute  superconductor. For larger densities bipolarons overlap and a new strongly correlated state may emerge.  We envision a few possibilities: either pairing correlations continue to dominate and $T_\mathrm{c}$ saturates or  grows (dashed line)  or competing effects suppress $T_\mathrm{c}$ (dotted line). Ultimately, at or near half filling, barring superconductivity at weak $\lambda$ in absence of nesting, $T_\mathrm{c}$  vanishes and, depending on the value of the electron-phonon coupling $\lambda$, a valence bond solid (VBS) or an antiferromagnet (AFM) emerges~\cite{BoSSH2D,CaiAFM,Assaad2D,ScaletterSSHU} (gray region).}
\label{fig:Fig4}
\vspace{-4mm}     
\end{figure}

\begin{figure*}
\centering
\vspace{-3mm}
\includegraphics[width=2\columnwidth]{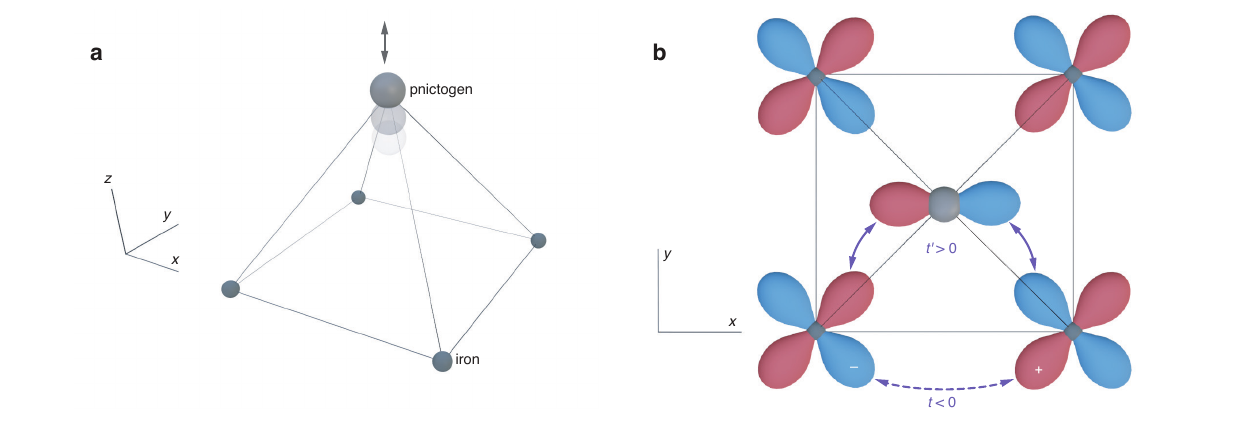}
\caption{{\bf Bond-Peierls coupling in the pnictides.}  A pnictogen atom sits at the apex of an octahedron with four iron atoms residing in the $x$-$y$ plane in the middle of the octahedron. {\bf a}. The phonon associated with the transverse motion of the pnictogen atom out of the $x$-$y$ plane in the $z$ direction causes fluctuations in the barrier for electronic tunneling between the iron atoms within the $x$-$y$ plane~\cite{HauleNatMater,PnictogenHeight}. {\bf b}. Interference pattern  along the bond connecting the iron atoms. Here there are two hopping processes: one resulting from direct overlap between $d_{xy}$ orbitals on neighboring iron atoms with amplitude $t<0$ (dashed double-headed arrow) and another involving a second-order process in which the $d_{xy}$ orbitals on iron atoms overlap with the $p_x$ orbital of the pnictogen atom (solid double-headed arrows), resulting in a hopping with amplitude $t'>0$. Destructive interference between these two processes results in a reduced net hopping along this pathway, e.g. in FeSe, see Appendix~\ref{Appendix:AppG} for more details.
}
\label{fig:Fig5}
\end{figure*}

\paragraph{Evolution with density of bipolaronic superconductivity.} Our estimates of $T_\mathrm{c}$ for bipolaronic superconductivity correspond to the largest electronic density for which bipolarons do not overlap.  At higher densities, studies  of  the emerging  strongly correlated state require new techniques capable of discerning between competing phases. We imagine at least two possibilities, depicted in Fig.~\ref{fig:Fig4}.  Either strong pairing correlations between the electrons~\cite{ModelQuarterFilling} result in saturation of $T_\mathrm{c}$ or even higher values of  $T_\mathrm{c}$ (dashed line), or competing effects resulting from the bipolaron overlap combined with the Hubbard repulsion and lattice reconstruction suppress $T_\mathrm{c}$ (dotted line).  In either scenario, the system eventually becomes non-superconducting at sufficiently large densities near or at half filling (unless unnested) and, depending on the value of $\lambda$, antiferromagnetism or valence-bond-solid charge order develops~\cite{BoSSH2D,CaiAFM,Assaad2D,ScaletterSSHU}.

\section{Outlook}\label{Outlook}

Obtaining high-transition-temperature superconductivity from Bose condensation of bipolarons has for many years been thought to be very difficult. Recent work~\cite{SousBipolaron} revealed conditions under which light bipolarons can arise, opening a door for phonon-mediated high-$T_\mathrm{c}$ superconductivity. This paper  uses an exact quantum Monte Carlo treatment of a precisely defined 2D square lattice model to demonstrate that bond electron-phonon coupling in fact gives rise to bipolaronic superconductivity with a transition temperature that is much higher than that obtained from the more extensively studied Holstein (density-coupled) bipolarons or  from  Migdal-Eliashberg theory of superconductivity out of a Fermi liquid. Perhaps more importantly, we find that $T_\mathrm{c}$ of the bond-coupled bipolaronic superconductor is enhanced by local Coulomb repulsion. The key ingredient in this bipolaronic mechanism that gives rise to high $T_\mathrm{c}$ is the combination of light mass and relatively small size of bipolarons. While the calculations reported here employ the bond-Peierls coupling in which the amplitude for an electron to hop between two sites is modulated by a phonon defined on the bond connecting the two sites, bipolaronic high-$T_\mathrm{c}$ superconductivity may also arise in the site-coupled Peierls (SSH) model~\cite{SousBipolaron} where the hopping is modulated by the relative distance between the atoms at the two ends of the bond, but an analysis including anharmonic couplings is required. These remarkable properties call for  consideration of the physics of Peierls electron-phonon coupling in quantum materials, and motivate further theoretical study of other physical situations that may give rise to small-size, light-mass bipolarons needed in order to support a state with high-$T_\mathrm{c}$ superconductivity. Models of potential interest include ones in which a phonon on a bond couples to two sites or a plaquette with either Holstein (e.g. Ref.~\cite{BreathingMode}) or Peierls coupling or both, or a situation in which $N$ phonons, each couples to $N$ electronic sites, see e.g. Ref.~\cite{KivelsonPhononLargeN}.

The bond-Peierls coupling studied here arises generically in materials where the orbitals of out-of-plane atoms mix with the bonding orbitals of in-plane atoms so that transverse fluctuations of the distance of the out-of-plane atoms give rise to modulation of the hopping. This physics may be operative in several families of materials, including the 90$\degree$-bonded~\cite{90bond,ZXNagaosaPeierlsCuprates} and corner-sharing~\cite{corner-sharing} perovskites. One particularly intriguing example occurs in the iron pnictides where electron transfer between two adjacent Fe ions can occur via a state on an intermediate pnictogen ion so that modulation of the pnictogen height strongly modulates particular hopping pathways~\cite{HauleNatMater,PnictogenHeight}, and where for certain geometries different pathways destructively interfere, producing a bond-Peierls coupling with a large coupling constant. Figure~\ref{fig:Fig5} demonstrates this scenario. See Appendix~\ref{Appendix:AppG} for more details and discussion. Although the model studied here is not directly applicable to the  pnictides, which are complex materials involving multi-orbital ``Hund's metal'' physics, it is intriguing to note that an ``extended $s$-wave'' state with some similarities to our bipolaron state has been proposed for its superconductivity. 

From a materials science perspective, it is worth highlighting that the high-$T_{\mathrm{c}}$ bipolaronic superconductivity becomes notably pronounced in the most ``quantal'' regime of $t/\Omega \sim 1$ (Fig.~\ref{fig:Fig1}) in which phonons and electrons are competitive energetically.  Here, $T_\mathrm{c}\sim 0.2 \Omega$, which could easily give rise to  a $T_{\mathrm{c}} \sim 70$K  for a typical value of phonon frequency $\sim 0.03$ eV  if and only if the unusual limit of $t \sim \Omega$ can be achieved.  For the material FeSe we expect the ratio of the nearest-neighbor hopping amplitude in the $x\;\mbox{-}\;y$ plane to the relevant out-of-plane phonon frequency to be $t/\Omega \sim 2\; \mbox{-}\;3$~\cite{HauleNatMater,ZXSHENFESE} (see Appendix~\ref{Appendix:AppG}), which is close to this ideal regime. In addition, structural engineering of a crystal's electronic stiffness~\cite{FunctionalTMOxides}, for example, by strain or moir\'e engineering~\cite{MoireFunctional} as found in twisted bilayer graphene~\cite{MAG}, can produce a reduction in $t$ without significantly changing $\Omega$,  perhaps realizing $t \sim \Omega$.   Alternatively, functional superatomic crystal engineering~\cite{SuperatomicMaterials}  may enable synthesis of crystals with light atoms on the bonds of the lattice, providing a path to high-$\Omega$ phonons, achieving $t \gtrsim  \Omega$ and an even higher value of $T_\mathrm{c}$. 

Important future research directions raised by our work include full theoretical characterization of the phenomenology of bipolaronic high-$T_{\mathrm{c}}$ superconductivity, extension of our results to the richer models describing the Hund's and multiorbital physics of materials such as the pnictides, and search for other bond-Peierls coupled compounds, thus opening a door to a new route to design principles of novel high-temperature superconductors.

\begin{acknowledgements}
We acknowledge useful discussions with F. Marsiglio. C.~Z. acknowledges support from the National Natural Science Foundation of China (NSFC) under Grant No. 12204173. J.~S., D.~R.~R. and A.~J.~M. acknowledge support from the National Science Foundation (NSF) Materials Research Science and Engineering Centers (MRSEC) program through Columbia University in the Center for Precision Assembly of Superstratic and Superatomic Solids under Grant No. DMR-1420634.  J.~S. acknowledges support from the Gordon and Betty Moore Foundation’s EPiQS Initiative through Grant GBMF8686 at Stanford University. M.~B. acknowledges support from the Natural Sciences and Engineering Research Council of Canada (NSERC), the Stewart Blusson Quantum Matter Institute (SBQMI) and the Max-Planck-UBC-UTokyo Center for Quantum Materials. N.~V.~P. and B.~V.~S. acknowledge support from the NSF under Grant No. DMR-2032077.  J.~S. also acknowledges the hospitality of the Center for Computational Quantum Physics (CCQ) at the Flatiron Institute. The Flatiron Institute is a division of the Simons Foundation.
\end{acknowledgements}

\appendix

\section{Conventions used for the definition of $\lambda$ in the different models}\label{Appendix:AppA}
In this appendix we detail the conventions used to define the dimensionless electron-phonon coupling constant $\lambda$ for the different models considered in the main text.

\begin{table*}
\renewcommand{\arraystretch}{1.5}
\centering
\caption{\label{tab:conventions}Definition of $\lambda$ for bipolarons and in Migdal-Eliashberg (ME) theory in the bond-Peierls (BP) and Holstein (H) models as employed in this work.}
\begin{tabular}{l | c  c}
\hline\hline
\phantom{Title} & \phantom{model} Bond-Peierls (BP) model \phantom{model} & \phantom{model}  Holstein (H) model \phantom{model}\\ \hline
Bipolaron  & $\frac{\alpha_\mathrm{BP}^2}{2\Omega t}$ & $\frac{\alpha_\mathrm{H}^2}{8\Omega t}$ \\
Migdal-Eliashberg (ME) theory & $\frac{8\alpha_\mathrm{BP}^2}{2\pi\Omega t}$ & $\frac{\alpha_\mathrm{H}^2}{2\pi\Omega t}$ \\
\hline\hline
\end{tabular}
\end{table*}

The electron-phonon coupling terms in the bond-Peierls (BP) and Holstein (H) models are
\begin{eqnarray}
\hat{\mathcal{V}}_\mathrm{BP} = \tilde{\alpha}_\mathrm{BP}\sum_{i,\sigma}^{}\Big\{\left( \hat{c}_{i,\sigma}^\dagger \hat{c}_{i+\hat{x},\sigma} + \mathrm{h.c.}\right)&\hat{X}_{i,i+\hat{x}}&
\nonumber \\ +\left( \hat{c}_{i,\sigma}^\dagger \hat{c}_{i+\hat{y},\sigma} +
\mathrm{h.c.}\right)&\hat{X}_{i,i+\hat{y}}&\Big\},
\label{Eq:BP}
\\
\hat{\mathcal{V}}_\mathrm{H} = \tilde{\alpha}_\mathrm{H}\sum_{i,\sigma}^{}\hat{c}_{i,\sigma}^\dagger \hat{c}_{i,\sigma} \hat{X}_i,&&
\label{Eq:HH}
\end{eqnarray}
where $\tilde{\alpha}=\alpha {\sqrt{2M\Omega}}$.  Our notation here differs slightly from that of the main text to clearly distinguish the BP and H models and to make explicit that there are two BP phonons in each unit cell of the square lattice, one associated with a bond in the $x$ direction and one with a bond in the $y$ direction.

In the polaronic limit, $\lambda$ is normally defined as the ratio of the typical polaron energy scale to the free electron energy scale, which for the BP model is $\lambda = \alpha_\mathrm{BP}^2/(2\Omega t)$ and for the H model is conventionally $\lambda = \alpha_\mathrm{H}^2/(4\Omega t)$~\cite{MacridinBipolaron}. However, in order to contrast $T_\mathrm{c}/\Omega$ for the bipolaronic superfluid in the two models on the same scale, we rescale $\lambda \rightarrow \lambda/2$ in the case of the H model so that it becomes $\lambda = \alpha_\mathrm{H}^2/(8\Omega t)$.  

In Migdal-Eliashberg (ME) theory of strong-coupling superconductivity out of a Fermi liquid, $\lambda$,  which we denote as $\lambda_\mathrm{ME}$, is conventionally defined via the Fermi surface mass enhancement:
\begin{eqnarray}
\left.\frac{m^\star}{m}\right\vert_\mathrm{FS} = 1 + \lambda_\mathrm{ME}.
\end{eqnarray}
Thus, $\lambda_\mathrm{ME}$ acquires a dependence on the density of states (see details in Appendix~\ref{Appendix:AppD}) and in the low electron density limit becomes $\lambda_\mathrm{ME} = 8\alpha_\mathrm{BP}^2/(2\pi\Omega t)$ in the BP model and $\lambda_\mathrm{ME} = \alpha_\mathrm{H}^2/(2\pi\Omega t)$ in the H model.

Table~\ref{tab:conventions} summarizes the conventions we use for the definition of $\lambda$ in the different models considered in this work.

\section{Quantum Monte Carlo calculations}\label{Appendix:AppB}

We employ a recently developed quantum Monte Carlo (QMC) approach based on a path-integral representation of the electronic sector combined with either a path-integral or a diagrammatic representation of the phononic sector.  The method and its details can be found in Ref.~\cite{QMCBondBipolaron}. Here we provide a brief overview of the approach, and details of the numerical simulations.

\subsection{Methodology}

We use Monte Carlo (MC) to stochastically sample the imaginary-time propagator ${\cal G}_{ba}(\tau) \equiv \bra{b} {e^{- \tau \hat{\cal H}}}\ket{a}$,
where $\tau$ is imaginary time, and $\ket{a}$ and $\ket{b}$ are any two-electron states in the singlet sector on a two-dimensional (2D) square lattice. We can evaluate the ground-state (GS) expectation value of any same- or different-time observable $\hat{O}$, defined as $\bar{O}_{\mathrm{GS}} \equiv \bra{\mathrm{GS}} \hat{O} \ket{\mathrm{GS}}$,  using Monte Carlo estimators:
\begin{equation}
\bar{O}_{\mathrm{GS}} = \frac{O_{ba} (\tau ) } { {\cal G}_{ba} (\tau )  }    \equiv \frac{\sum_{ab} W_{ab} O_{ba} (\tau ) } { \sum_{ab} W_{ab} {\cal G}_{ba} (\tau )} 
\label{MCestimators},
\end{equation}
where $O_{ba}(\tau)  =  \bra{b}{e^{- (\tau /2) \hat{\cal{H}}} \hat{O}  e^{- (\tau /2) \hat{\cal{H}}}}\ket{a}$ and $W_{ab}$ are MC weights.

The approach to MC sampling is based on the general scheme proposed in Refs.~\cite{Worm1,Worm2}. We write $\hat{\cal{H}} = \hat{h} + \hat{V}$, where $\hat{h}$ and $\hat{V}$ are  the diagonal and off-diagonal parts of the
Hamiltonian with respect to a basis ${\cal B} = \{ \ket{\alpha}\}$:  $\hat{h} \ket{\alpha} = E_{\alpha}  \ket{\alpha}$, 
$\hat{V} = \sum_{\alpha \beta} V_{\beta \alpha} \ket{\beta} \bra{\alpha }$,  and $\bra{\alpha } \hat{V} \ket{\alpha} =0$. For the model in Eqs.~\eqref{Eq:H}, \eqref{Eq:Veph}, the basis ${\cal B}$ corresponds to site Fock states for the electrons and bond Fock states for the phonons. Using the interaction representation for the evolution operator in imaginary time:
$e^{-\tau \hat{\cal H}}  = e^{-\tau \hat{h}} \hat{\sigma} (\tau) $,
and expanding $\hat{\sigma} (\tau)$ one finds~\cite{Worm1,Worm2}:
\begin{widetext}
\begin{eqnarray}
 \sigma_{\beta \alpha}(\tau)  =
\delta_{\alpha \beta} \, -  \int_0^{\tau}
d \tau_1 \: V_{\beta \alpha} \, e^{\tau_1 E_{\beta \alpha}}  + \sum_{\gamma_1} \!  \int_{0}^{\tau}  \! \! \!  d \tau_2 \! \!
                       \int_0^{\tau_2}  \! \! \!  d \tau_1
V_{\beta \gamma_1 }  e^{\tau_2 E_{\beta    \gamma_1}}
V_{\gamma_1 \alpha} e^{\tau_1 E_{\gamma_1 \alpha}  } + \ldots  ,  \qquad
\label{sigma}
\end{eqnarray}
\end{widetext}
where $E_{\beta \alpha} = E_{\beta} - E_{\alpha}$.   In this  representation, there are three types of the elementary processes: 1. bare electron hopping, 2. electron hopping assisted by phonon creation, and 3. electron hopping assisted by phonon annihilation.  Phonons can also be treated diagrammatically within the same expansion whilst maintaining a path-integral representation only for the electronic sector. The MC scheme employed  is based on the statistical interpretation of the right-hand side of Eq.~\eqref{sigma} as an
average over an ensemble of graphs, in which each graph represents a string in space-time coordinates characterized by the number and types of kinks $V_{\gamma_{i+1} \gamma_{i}}$. For the model in Eqs.~\eqref{Eq:H}, \eqref{Eq:Veph}, the choice of basis ${\cal B}$ ensures that $-V_{\beta \alpha}$ are non-negative, real numbers, and thus graphs are sampled according to non-negative weights given by the values of the corresponding
integrands in Eq.~\eqref{sigma}, rendering this  a sign-problem-free MC approach. The sign problem is present for the site-Peierls model because the coupling involves the difference between phonon displacement operators on different sites, which means that the sign of a subset of the $-V_{\beta \alpha}$ factors in Eq.~\eqref{sigma} will be negative. In contrast, in the bond-Peierls model, the sign of the coupling can be always gauged away and therefore the sign problem is absent. We use the following updates in the MC scheme: we stochastically 1. add and remove the last bare-hopping kink using a pair of complementary updates \cite{Worm1,Worm2}, 2. switch between the three types of the hopping terms, and 3. sample the length of the last time interval separating the last kink from the state $\bra{b}$. This scheme yields states $\bra{b}$ that admit any allowed configuration of excited phonon modes, and, as a result, ensures ergodicity.

\begin{figure*}
\centering
\vspace{-3mm}
\includegraphics[width=1.85\columnwidth]{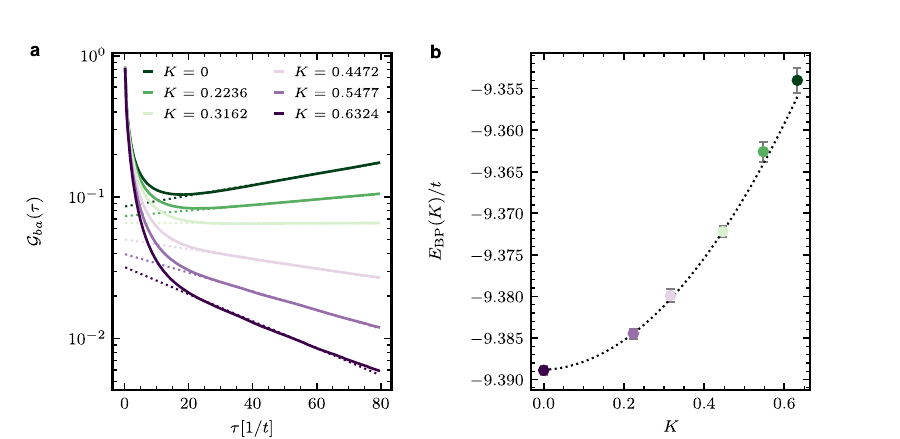}
\caption{{\bf Computation of the bipolaron mass using a QMC approach based on a path-integral representation of the electrons combined with a diagrammatic treatment of the phonons.} {\bf a}. Green function ${\cal G}_{ba} (\tau )$ as a function of imaginary time $\tau$ in different bipolaron momentum ($K$) sectors (solid lines) and fits to the long-$\tau$ asymptotic behavior of Eq.~\eqref{eq:EqGSPIQMC} (dashed lines). {\bf b}. The bipolaron dispersion $E_\mathrm{BP} (K)$ constructed from $E_\mathrm{GS}$ in the different $K$ sectors obtained in {\bf a} from the fits of ${\cal G}_{ba} (\tau )$ in the large-$\tau$ limit to the asymptotic form in Eq.~\eqref{eq:EqGSPIQMC}. We compute the bipolaron mass by fitting the dispersion. For this data set, we find the following fitting function: $E_\mathrm{BP}(K) = -9.38883 +0.00188511 K + 0.0791534 K^2$, which yields $m^{\! \star}_\mathrm{BP}/m_0 =  6.33\pm 1.0$. Error bars represent statistical errors in QMC simulations corresponding to one standard deviation. Results shown in this figure are for the BP bipolarons at $\lambda = 0.5$, $t/\Omega = 10/3$ and $U/t=8$.}
\label{fig:FigA1}
\vspace{-1mm}
\end{figure*}

\begin{figure}[!b]
\centering
\includegraphics[width=0.92\columnwidth]{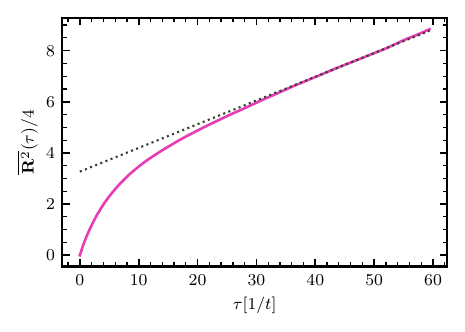}\hspace{2mm}
\caption{{\bf Computation of the bipolaron mass using a QMC approach based on a path-integral representation of both the electrons and the phonons.} Mean-square fluctuations of the relative displacement $\overline{{\bf R}^2}(\tau)$ as a function of imaginary time $\tau$ (solid line) and fit to the long-$\tau$ asymptotic behavior of Eq.~\eqref{R2} (dashed line) which we use to obtain the bipolaron mass. For this data set, we find the following fitting function: $\overline{{\bf R}^2}(\tau) /4 = 3.26899 + 0.0927293 \tau$, which yields $m^{\! \star}_\mathrm{BP}/m_0 =  5.4\pm 0.3$. Results shown in this figure are for the BP bipolarons at $\lambda = 0.5$, $t/\Omega = 10/3$ and $U/t=8$.}
\vspace{-1mm}   
\label{fig:FigA2}
\end{figure}

\subsection{Computation of bipolaron properties}\label{Appendix:AppB2}
First, we note that the imaginary-time dependence of ${\cal G}_{ba}(\tau)$
contains direct information about the ground-state energy $E_\mathrm{GS}$ within a given momentum symmetry sector of the Hilbert space:
\begin{equation}
{\cal G}_{ba}(\tau)  \mathop  {\longrightarrow} \limits_{\tau \to \infty}
\bra{b}\ket{\mathrm{GS}}\bra{\mathrm{GS}}\ket{a} e^{- \tau E_{\mathrm{GS}}},
\label{eq:EqGSPIQMC}
\end{equation}
and thus we can use this relation to  extract  $E_\mathrm{GS}$ in the $\tau \to \infty$ limit for a given momentum symmetry sector, see Fig.~\ref{fig:FigA1}. Furthermore, in the center-of-mass coordinate representation of the two-electron states $\ket{a}$ and $\ket{b}$, the propagator ${\cal G}_{ba} (\tau, \mathbf{R})$ which depends on the relative center-of-mass displacement $\mathbf{R}=\mathbf{R}_b - \mathbf{R}_a$, assumes a universal form given by the propagator of a free particle whose effective mass is given by the bipolaron mass $m^{\! \star}_\mathrm{BP}$ (see, for example, Ref.~\cite{massQMCuniversal}), which in 2D takes the form:
\begin{equation}
{\cal G}_{ba} (\tau , {\bf R})   \to  {A_{ba}   e^{-E_\mathrm{GS} \tau}  \over \tau} \, e^{-{m^{\! \star}_\mathrm{BP} {\bf R}^2 \over 2 \tau}},
\label{Gauss}    
\end{equation}
where $A_{ba}$ is a non-universal coefficient which depends on the choice of states $\ket{a}$ and $\ket{b}$.
We can thus extract $m^{\! \star}_\mathrm{BP}$, by averaging over the states $\ket{a}$ and $\ket{b}$, from the mean-square fluctuations of the center-of-mass displacement in the large-$\tau$ limit:
\begin{equation}
\overline{{\bf R}^2} (\tau)  = \frac{\sum_{ab} W_{ab} {\cal G}_{ba} (\tau,{\bf R}) {\bf R}^2 } {\sum_{ab} W_{ab} {\cal G}_{ba} (\tau, {\bf R})} \mathop  {\longrightarrow}\limits_{\tau \to \infty} \frac{2}{m^{\! \star}_\mathrm{BP}}\tau.
\label{R2}
\end{equation}
$\overline{{\bf R}^2} (\tau)$ ultimately saturates to a straight line at sufficiently large $\tau$ leading to an accurate estimate of the effective mass, see Fig.~\ref{fig:FigA2}.

\begin{figure}[!t]
\centering
\includegraphics[width=0.92\columnwidth]{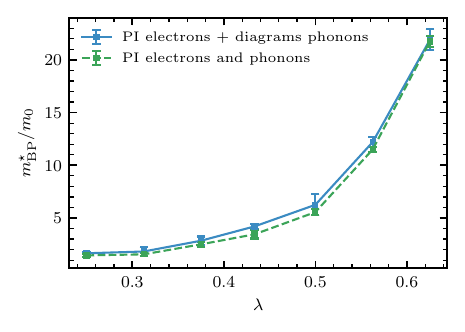}\hspace{2mm}
\caption{{\bf Bipolaron mass $m^{\! \star}_\mathrm{BP}$ in units of the mass of two free electrons $m_0 = 2m_e = 1/t$ obtained using a QMC approach based on a path-integral representation of the electrons combined with a diagrammatic treatment of the phonons (solid line) versus that obtained using a QMC approach based on a path-integral representation of both the electrons and the phonons (dashed line).} Results shown in this figure are for the BP bipolarons at $t/\Omega = 10/3$ and $U/t=8$.}
\vspace{-1mm}   
\label{fig:FigA3}
\end{figure}

\begin{figure*}
\centering
\includegraphics[width=1.8\columnwidth]{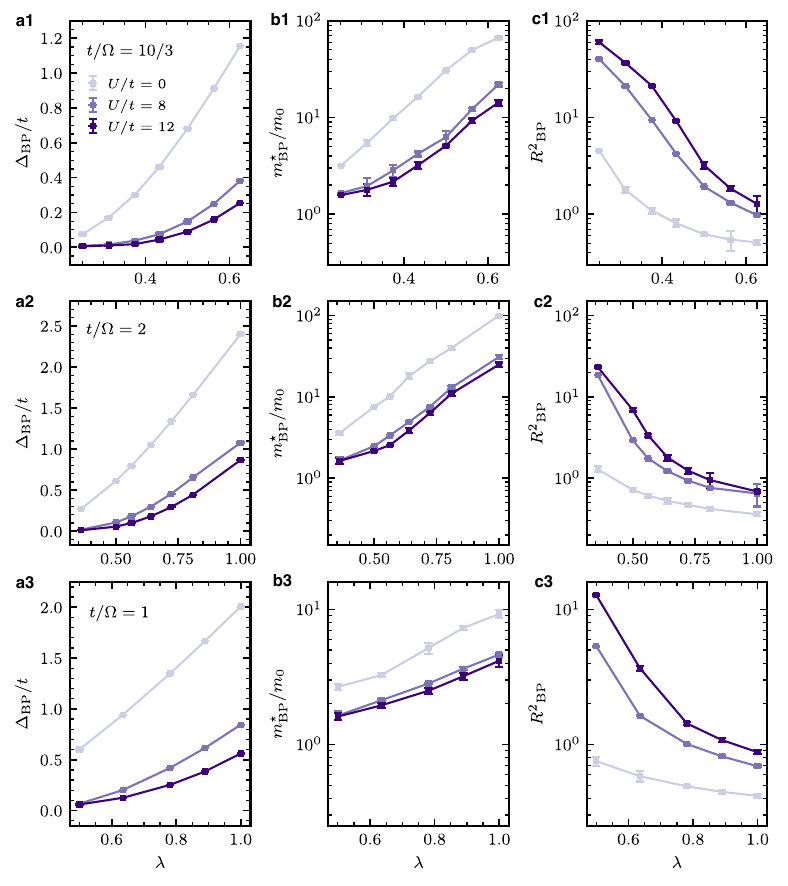}
\vspace{-4mm}
\caption{{\bf Bipolaron properties in the bond-Peierls model.} Bipolaron properties computed from QMC calculations performed on Eq.~\eqref{Eq:H} at various values of the adiabaticity parameter  $t/\Omega$ (different rows) as a function of the electron-phonon coupling  $\lambda = \alpha^2/(2\Omega t)$ for different onsite Hubbard $U$ (in units of the electron hopping $t$): {\bf a1},{\bf a2}, {\bf a3}. Bipolaron binding energy $\Delta_{\mathrm{BP}}$ in units of the electron hopping $t$, {\bf b1},{\bf b2}, {\bf b3}. Bipolaron effective mass $m^{\! \star}_\mathrm{BP}$ in units of the mass of two free electrons $m_0 = 2m_e = 1/t$, and {\bf c1},{\bf c2}, {\bf c3}. Bipolaron mean squared-radius $R^2_{\mathrm{BP}}$. Error bars represent statistical errors in QMC simulations corresponding to one standard deviation.}\label{fig:FigA4}
\vspace{-4mm}
\end{figure*}

\begin{figure*}
\centering
\includegraphics[width=1.778\columnwidth]{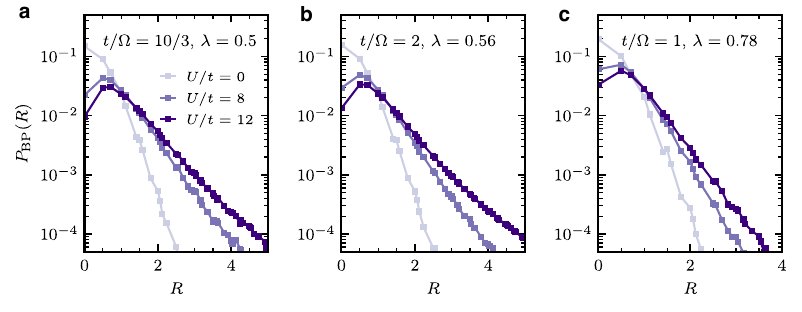}
\vspace{-4mm}
\caption{{\bf Bipolaron radial size probability density distribution $P_\mathrm{BP}(R)$ in the bond-Peierls model.} $P_\mathrm{BP}(R)$ computed from QMC calculations performed on Eq.~\eqref{Eq:H} at various values of the adiabaticity parameter $t/\Omega$ for the value of $\lambda = \alpha^2/(2\Omega t)$ which maximizes $T_\mathrm{c}$  at onsite  Hubbard repulsion $U/t=0, 8, 12$; see Fig.~\ref{fig:Fig1}. Error bars in $P_\mathrm{BP}(R)$ correspond to statistical errors smaller than the symbol size and therefore are not shown.}\label{fig:FigA5}
\vspace{-2mm}
\end{figure*}

We can thus compute the bipolaron mass either by constructing the bipolaron dispersion $E_\mathrm{BP}(K)$ as a function of the bipolaron momentum $K$ from the asymptotic behavior of Eq.~\eqref{eq:EqGSPIQMC} as shown in Fig.~\ref{fig:FigA1} or directly from the asymptotic behavior of $\overline{{\bf R}^2}(\tau)$ in Eq.~\eqref{R2} as shown in Fig.~\ref{fig:FigA2}. We use a QMC approach based on a path-integral representation of the electrons combined with a diagrammatic treatment of the phonons to simulate the asymptotic behavior of Eq.~\eqref{eq:EqGSPIQMC}, and a QMC approach based on a path-integral representation of both the electrons and the phonons to simulate asymptotic behavior of Eq.~\eqref{R2}.  Estimates of the mass obtained using these two approaches agree within the error bars, see Fig.~\ref{fig:FigA3}. Importantly, the dispersion  we obtain in our simulations retains a parabolic form for all energies lower than $T_\mathrm{c}$ even when $K$ becomes on the order of the inverse size of the bipolaron, justifying the use of a continuous space description.

To compute the bipolaron mean squared-radius $R^2_{\mathrm{BP}}$ we use  Monte Carlo estimators (Eq.~\eqref{MCestimators}) to evaluate  the probability distribution $P(R_{12})$ of finding the two electrons at a distance $R_{12}$ from their center-of-mass position within a  bound bipolaron, from which we compute $R^2_{\mathrm{BP}} \equiv \langle R_{12}^2 \rangle = \sum_{R_{12}} R_{12}^2 P(R_{12})$. (Note that $R^2_{\mathrm{BP}}$ and $\overline{{\bf R}^2} (\tau)$ of the two-electron state are unrelated; the former is the square of the relative distance measured from the center-of-mass coordinate, i.e. half the distance between the two electrons, while the latter is the square of the center-of-mass displacement---during  imaginary-time evolution---averaged  over worldline configurations.)

We use the two QMC approaches to simulate the behavior of two electrons in the singlet sector of the model (Eqs.~\eqref{Eq:H}, \eqref{Eq:Veph}) on a 2D square lattice with linear size $L=128$ sites. The accuracy of the QMC results is controlled by the maximum value of $\tau$, $\tau_{\rm max}$ which determines the accuracy of the projection of the propagation onto the ground state, with numerically exact results recovered in the limit $\tau_{\rm max} \rightarrow \infty$. All results presented in this work are converged with respect to $\tau_{\rm max}$.  Our calculations of the mass and size of bipolarons in the ground state allow us to estimate $T_\mathrm{c}$ reliably at temperatures below the binding gap, i.e., so long as $T_\mathrm{c}\leq \Delta_\mathrm{BP}$.  Error bars shown in the figures represent statistical errors in QMC simulations corresponding to one standard deviation, and, when applicable, account for errors in fitting.

\section{Supplementary results on superfluidity of bond-Peierls bipolarons}\label{Appendix:AppC1}

\begin{figure*}
\centering
\vspace{-3mm}
\hspace{3mm}\includegraphics[width=1.85\columnwidth]{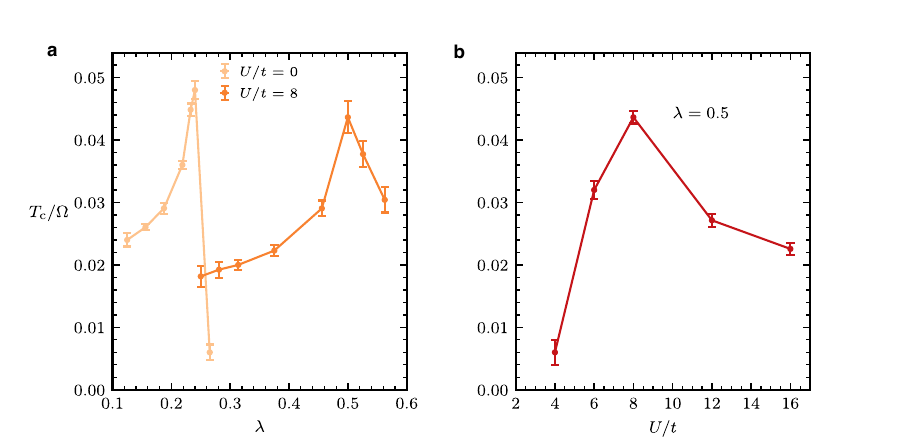}
\caption{{\bf Bipolaronic superconductivity in the Holstein model.} $T_\mathrm{c}/\Omega$ at adiabaticity parameter $t/\Omega = 2$ as a function of  $\lambda = \alpha^2/(8\Omega t)$ for an onsite Hubbard $U/t = 0$ and $8$ ({\bf a}) and as a function of $U/t$ for $\lambda = 0.5$ ({\bf b}). Error bars represent statistical errors in QMC simulations corresponding to one standard deviation.}
\vspace{0mm}   
\label{fig:FigA6}
\end{figure*}

In the main text, in Fig.~\ref{fig:Fig1}, we show $T_\mathrm{c}$ of a superfluid of bond-Peierls bipolarons at $U/t=8$ at various values of the adiabaticity parameter $t/\Omega$. The behavior of the mass and size of the bipolaron determines the value of $T_\mathrm{c}$ as can be seen from Eq.~\eqref{Eq:Tc}. Figure~\ref{fig:Fig3} provides information about the bipolaron mass and size at $t/\Omega = 10/3$.  Here we present  in Figs.~\ref{fig:FigA4}, \ref{fig:FigA5} additional results on the properties of the bipolaron for various values of $t/\Omega$.

Figure~\ref{fig:FigA4} shows that the overall trend of $\Delta_\mathrm{BP}$, $m^*_\mathrm{BP}$ and $R_\mathrm{BP}^2$ with $\lambda$ shifts to larger values of $\lambda $ as $t/\Omega$ decreases, but attains a qualitatively similar profile. Similarly, the spatial structure of the bipolaron $P_\mathrm{BP}(R)$ at the optimal $\lambda$ that maximizes $T_\mathrm{c}$ appears to not depend strongly on the value of $t/\Omega$, see Fig.~\ref{fig:FigA5}. From Fig.~\ref{fig:FigA4} we see that upon increasing $\lambda$, both $m^*_\mathrm{BP}$ increases and $R^2_\mathrm{BP}$ decreases in a fashion in which there exists an optimal $\lambda$ (e.g. for $t/\Omega = 10/3$, the optimal $\lambda$ is  $\sim 0.5$) for which $m^*_\mathrm{BP}$ is not too large yet $R^2_\mathrm{BP}$ is relatively small, leading to a maximum in the $T_\mathrm{c}$ curve, see Fig.~\ref{fig:Fig1}.

\begin{figure*}
\centering
\vspace{0mm}
\hspace{-2mm}\includegraphics[width=1.85\columnwidth]{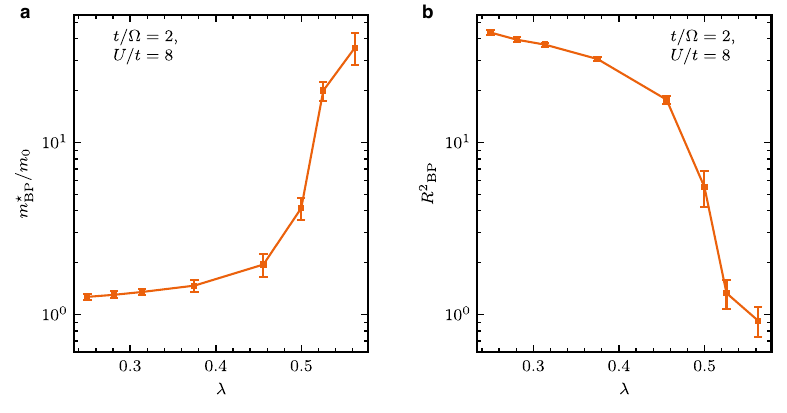}
\caption{{\bf Bipolaron properties in the Holstein model.}
Bipolaron properties computed from QMC calculations performed on Eq.~\eqref{Eq:HH} at adiabaticity parameter $t/\Omega = 2$ as a function of the electron-phonon coupling  $\lambda = \alpha^2/(8\Omega t)$ for an onsite Hubbard $U/t=8$: {\bf a}. Bipolaron effective mass $m^{\! \star}_\mathrm{BP}$ in units of the mass of two free electrons $m_0 = 2m_e = 1/t$, and {\bf b}. Bipolaron mean squared-radius $R^2_{\mathrm{BP}}$. Error bars represent statistical errors in QMC simulations corresponding to one standard deviation.}
\vspace{-1mm}   
\label{fig:FigA7}
\end{figure*}

\section{Supplementary results on superfluidity of of Holstein bipolarons}\label{Appendix:AppC2}

In the main text, in Fig.~\ref{fig:Fig1}, we show $T_\mathrm{c}$ of a superfluid of Holstein bipolarons at $U/t=8$.  Here we present additional results on the behavior of $T_\mathrm{c}$ of a superfluid of Holstein bipolarons computed from Eq.~\eqref{Eq:Tc} using the same methodology as for the bond-Peierls bipolarons, and provide details on the behavior of the mass and radial size of the bipolaron.

In Fig.~\ref{fig:FigA6}, we show the behavior of $T_\mathrm{c}/\Omega$ for Holstein bipolarons at $t/\Omega = 2$ as a function of $\lambda$ for $U/t=0$ and $8$ (Fig.~\ref{fig:FigA6}{\bf a}), and as a function of $U/t$ for $\lambda = 0.5$ (Fig.~\ref{fig:FigA6}{\bf b}).  $T_\mathrm{c}/\Omega$ never exceeds $\sim 0.05$.  Increasing $\lambda$ past an optimal but small value leads to rapid bipolaron mass enhancement (see below) and suppression of $T_\mathrm{c}$. Increasing $U/t$ leads to a decrease in the mass, but the binding energy drops very rapidly and the radius correspondingly increases, limiting $T_\mathrm{c}$ to even smaller values. This behavior can be seen clearly in Fig.~\ref{fig:FigA7} which shows that $T_\mathrm{c}$ is much smaller in the Holstein model than in the bond-Peierls model because the Holstein bipolarons are always very heavy and small when strongly bound.  They can become lighter only when  weakly bound which also results in large $R^2_\mathrm{BP}$, once again producing a small $T_\mathrm{c}$.

\section{Migdal-Eliashberg theory calculations}\label{Appendix:AppD}

This appendix summarizes results obtained in the Migdal-Eliashberg (ME) approximation for the Holstein (H) and bond-Peierls (BP) models considered in the main text. The approximation and computations are standard, although the implications of the momentum-dependent electron-phonon coupling in the BP model have not previously been discussed. The purpose of the appendix is to  make precise  the comparison given  in the main text  of  the transition temperature $T_\mathrm{c}$ computed within ME theory by presenting the  specifics of the calculations. It is important to emphasize that in the literature on the limits on $T_\mathrm{c}$ in real compounds~\cite{McMillanFormula,CohenBounds} the  physics of the limits on $T_\mathrm{c}$ relies primarily on considerations of the maximum physically attainable electron-phonon coupling strength in realistic models in which the electron-phonon interaction is computed from microscopics in a theory of electrons and ions coupled by the physical Coulomb interactions. Here we focus on the properties of the model systems discussed in the main text, with the Hubbard $U$ set to zero. The results for the H model presented here are consistent with recent work of Esterlis and collaborators~\cite{TcBoundKivelson}.

Contact with conventional ME theory is more clearly made in momentum ($k$) space, where the H (Eq.~\eqref{Eq:HH}) and BP (Eq.~\eqref{Eq:BP}) couplings are
\begin{eqnarray}
\hat{\mathcal{V}}_\mathrm{H} &=& \tilde{\alpha}_\mathrm{H}\sum_{k,q,\sigma}^{}\left( \hat{c}_{k-\frac{q}{2},\sigma}^\dagger \hat{c}_{k+\frac{q}{2},\sigma} \right)\hat{X}_q,
\label{Eq:Hk}
\\
\hat{\mathcal{V}}_\mathrm{BP} &=& \tilde{\alpha}_\mathrm{BP}\sum_{a=\pm }\sum_{k,q,\sigma}^{}\left( \hat{c}_{k-\frac{q}{2},\sigma}^\dagger \hat{c}_{k+\frac{q}{2},\sigma} \right)\Lambda_a(k_x,k_y)\hat{X}_{q,a}. \nonumber \\
\label{Eq:BPk}
\end{eqnarray}
We have rewritten the BP coupling in terms of  the phonon operators:
\begin{equation}
\hat{X}_{q,\pm}=\frac{\hat{X}_{q,x}\pm \hat{X}_{q,y}}{\sqrt{2}},
\label{bpmdef}
\end{equation}
with couplings
\begin{equation}
    \Lambda_\pm (k_x, k_y)=\sqrt{2}\left(\cos k_x\pm \cos k_y\right).
    \label{eq:Lambdapmdef}
    \end{equation}

\subsection{Phonon stiffness and stability limits}
The ME theory of these models proceeds by first computing a physical phonon stiffness given by the difference of the bare phonon stiffness denoted by $K$ and $\tilde{\alpha}^2$ multiplied by the zero frequency limit of an appropriate electron correlation function. (In the adiabatic limit $\Omega_0\ll E_F$, the frequency dependence of the correlator is negligible; in other words, the phonon mass is not renormalized, and the correlator may be computed using the bare electron Green functions. Here $\Omega_0$ is the bare phonon frequency, which is the same as $\Omega$ in the main text.)

In the H model the relevant correlator is the electron density-density correlation function and we have
\begin{equation}
    K_\mathrm{H}(q)=K\left(1-\frac{\tilde{\alpha}^2}{K}\chi_{\rho\rho}(q,0)\right).
    \label{eq:KH}
\end{equation}
Since a positive phonon stiffness is required for stability of the oscillator, the maximum coupling is bounded (in the ME approximation) by $\tilde{\alpha}^2<\mathrm{max}_q\frac{K}{\chi_{\rho\rho}(q,0)}$. Standard density functional theory (DFT) computations of phonon frequencies include (within the approximations of DFT and within the adiabatic limit) renormalization of the phonon frequency, i.e. a material crystal structure is by construction stable (within the DFT approximation). Esterlis and collaborators~\cite{MEBreakdownKivelson, TcBoundKivelson} studied the stability issue using numerical methods which allowed them to go beyond the ME approximation; the main focus of their work was electron densities near half filling where the susceptibility has a substantial density dependence and density waves provide a competing ground state. A qualitative result of their work is that while the ME approximation is not accurate near the stability limit, the estimate $\tilde{\alpha}^2<\mathrm{max}_q\frac{K}{\chi_{\rho\rho}(q,0)}$ still provides a reasonable bound.

\begin{figure}[!t]
\centering
\includegraphics[width=0.92\columnwidth]{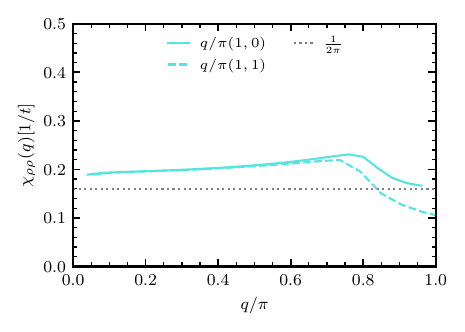} \vspace{-2mm}   
\caption{Static density-density correlation function $\chi_{\rho \rho} (q) \equiv \chi_{\rho \rho} (q,0)$ of non-interacting electrons on a 2D square lattice tight-binding model with dispersion $\varepsilon_k=-2t(\cos k_x+\cos k_y)$ at carrier concentration $n=0.25$ carrier per site as function of momentum ($q$) along direction $(1,0)$ (solid line) and $(1,1)$ (dashed line), with $n\rightarrow 0$ value of density of states $1/(2\pi t)$ (dotted line).}
\vspace{-3mm}   
\label{fig:chiholstein}
\end{figure}

\begin{figure*}
\vspace{4mm}
\centering
\includegraphics[width=2\columnwidth]{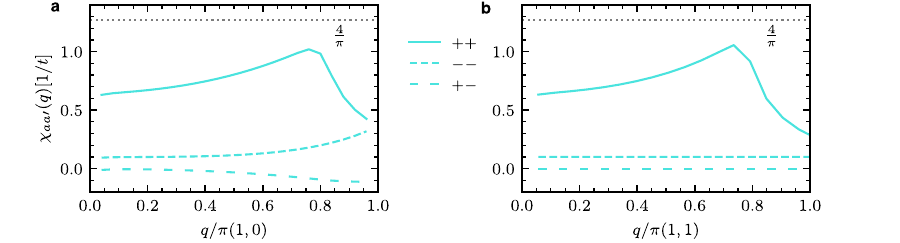} 
\caption{Static $(\cos k_x \pm \cos k_y)(\cos k_x\pm \cos k_y)$ correlation functions $\chi_{++}(q)$ (solid line), $\chi_{--}(q)$ (dashed line) and $\chi_{+-}(q)$ (spaced dashed line) (here $\chi_{aa\prime} (q) \equiv \chi_{a a\prime} (q,0)$)
of non-interacting electrons on a 2D square lattice tight binding model with dispersion $\varepsilon_k=-2t(\cos k_x+\cos k_y)$ at carrier concentration $n=0.25$ carrier per site as a function of momentum ($q$) along direction $(1,0)$ ({\bf a}) and $(1,1)$ ({\bf b}), with $n\rightarrow 0$ value of $\chi_{++}(q) = 4/(\pi t)$ (dotted line).}
\label{fig:chiBP}
\vspace{-3mm}
\end{figure*}

Our interest here is specifically in the ME approximation and in the low density limit. For the electronic model  dispersion used here $\varepsilon_k=-2t\left(\cos k_x+\cos k_y\right)$, at low electronic densities (say, $n\equiv\langle \hat{n}_i \rangle \lesssim 0.25$ electron per site where $\hat{n}_i = \hat{n}_{i,\uparrow} + \hat{n}_{i,\downarrow}$) the susceptibility is essentially momentum independent in the range $0\leq q \leq 2k_\mathrm{F}$, and is equal to the bare fermion density of states (summed over spin) $N_0$, which is weakly density dependent and tends to   $1/(2\pi t)$ as the density goes to zero, see Fig.~\ref{fig:chiholstein} where we show the static limit of the bare electron density-density correlation function computed for $n=0.25$ carrier per site.  Thus, in the $0$ - $2k_\mathrm{F}$ wave vector range the {\em physical} phonon frequency $\Omega_\mathrm{phys}$ is, to a good approximation, constant and given by
\begin{equation}
    \Omega_\mathrm{phys}=\Omega_0\sqrt{1-2\lambda^0_\mathrm{H;ME}},
    \label{eq:Omegaphys}
\end{equation}
where the bare ME coupling constant is
\begin{equation}
    \lambda_\mathrm{H;ME}^0=\frac{\alpha_\mathrm{H}^2}{\Omega_0}N_0\approx \frac{\alpha_\mathrm{H}^2}{2\pi \Omega_0 t},
\end{equation}
and the second, approximate, equality becomes exact  in the low density limit. Thus, the stability limit of the Holstein  model in the ME approximation (see Eq.~\eqref{eq:KH}) at low densities is $\lambda_\mathrm{H;ME}^0=1/2$. Note that the $\lambda$ defined for Holstein bipolarons differs from this definition, see Table~\ref{tab:conventions}.

In the BP model  there are two phonon modes per unit cell of the square lattice and the couplings are momentum-dependent. In the $\pm$ basis of Eq.~\eqref{Eq:BPk}  Eq.~\eqref{eq:KH} becomes
\begin{equation}
\bfit{K}_\mathrm{BP}(q)=K\left(\mathbf{1}-\frac{\tilde{\alpha}_\mathrm{BP}^2}{K}\left(\begin{array}{cc}\chi_{++}(q,0) & \chi_{+-}(q,0) \\\chi_{+-}(q,0) & \chi_{--}(q,0)\end{array}\right) \right),
\label{eq:KBP}
\end{equation}
where $\chi_{\pm\pm}$ is the $\langle \Lambda_{\pm}\Lambda_\pm \rangle$ correlator and $\chi_{+-}$ is the $\langle \Lambda_+ \Lambda_- \rangle$ correlator (see Eq.~\eqref{eq:Lambdapmdef}). Figure~\ref{fig:chiBP} shows numerical calculation of these correlators. The cross correlator $\chi_{+-}(q,0)$ is in general very small; we neglect it here. In the very low density limit ($n\lesssim 0.05/\mathrm{site}$), $\chi_{++}(q,0)=8\chi_{\rho,\rho}=\frac{4}{\pi t}$ and $\chi_{--}(q,0)\approx 0$. For moderately low densities (e.g. $n=0.25/\mathrm{site}$) $\chi_{--}(q,0)$ remains small relative to $\chi_{++}(q,0)$ but $\chi_{++}(q,0)$ acquires non-negligible momentum dependence with the largest value being at $q=2k_\mathrm{F}$. For larger $n$, $\chi_{--}$ can become larger than $\chi_{++}$ and in fact sets the limit of stability. The  momentum dependence of the electron-phonon coupling leads to some ambiguity in the definition of $\lambda_\mathrm{BP;ME}^0$. Here we adopt a definition appropriate to the very low density limit, writing
\begin{equation}
    {\Omega_\mathrm{phys}}_{\pm}(q)=\Omega_0\sqrt{1-2\lambda_\mathrm{BP;ME}^0\left(\frac{\pi t}{4}\chi_{\pm}(q,0)\right)},
\end{equation}
with
\begin{equation}
\lambda^0_\mathrm{BP;ME}=\frac{4\alpha_\mathrm{BP}^2}{\pi \Omega_0 t},
\end{equation}
where $\chi_\pm (q,0) \equiv \chi_{\pm \pm} (q,0)$.  The $\lambda$ defined for BP bipolarons in the main text is a factor of $\pi/8$ smaller than  $\lambda^0_\mathrm{BP;ME}$, see also Table~\ref{tab:conventions}.  The factor $\frac{\pi t}{4}\chi_{+}(q,0)$ becomes $1$, independent of $q$ in the very low density limit, and the $-$ mode decouples, so  the physics becomes identical to the H model apart from the relation between $\alpha$ and $\lambda$. At larger $n\sim 0.25$ the $-$ mode still decouples but the factor is less than $1$ at all $q$ and is $q$ dependent, being largest at $q=2k_\mathrm{F}$. At larger densities the $-$ mode does not decouple.

\subsection{Electron self energy}

The ME approximation to the electron self energy $\Sigma$ (making use of the adiabatic limit, which allows us to average the self energy over all wave vectors on  the Fermi surface) gives
\begin{widetext}
\begin{equation}
\Sigma(\omega)=T\sum_{\omega^\prime}\mathbf{\mathcal{G}}(\omega^\prime)\frac{\int\frac{d^2k}{(2\pi)^2}\delta\left(\varepsilon_{k}\right)\int\frac{d^2k^\prime}{(2\pi)^2}\delta\left(\varepsilon_{k^\prime}\right)\sum_a\pi\alpha^2_{a;k,k^\prime}D_a(k-k^\prime,\omega-\omega^\prime)}{\int\frac{d^2k}{(2\pi)^2}\delta\left(\varepsilon_{k}\right)},
\label{eq:sigmaME1}
\end{equation}
\end{widetext}
where the phonon propagator for mode $a$ is
\begin{equation}
D_a(q,\nu)=\frac{2\Omega_a(q)}{\nu^2+\Omega_a^2(q)}
\label{D}
\end{equation}
and the Fermi surface-projected electron Green function is 
\begin{equation}
\mathbf{\mathcal{G}}(\omega_\mathrm{n})=\frac{i\omega_\mathrm{n}\mathbf{1}+\tau_1W(\omega_\mathrm{n})}{\sqrt{(\omega_nZ(\omega_\mathrm{n}))^2+W(\omega_\mathrm{n})^2}},
\label{GQCdef}
\end{equation}
where $\nu$ and $\omega_\mathrm{n}$
are Matsubara frequencies for bosons and fermions, respectively, and $\tau_1$ in the first of the $\mathrm{SU(2)}$ Pauli matrices.   Here the normal component of the self energy is $\Sigma_\mathrm{n}=i\omega_\mathrm{n}(1-Z(\omega_\mathrm{n}))$ (where $Z(\omega_\mathrm{n})$ is known as the $Z$ factor) and the anomalous component is $W(\omega_\mathrm{n})$. In the circular Fermi surface approximation which is reasonably accurate for $n\leq 0.25$  we have $\varepsilon_k \approx-4t+k^2/(2m)$ with $m=\frac{1}{2t}$ and the self energy equation becomes
\begin{widetext}
\begin{equation}
\Sigma(\omega)=\frac{m}{2\pi}T\sum_{\omega^\prime}\mathbf{\mathcal{G}}(\omega^\prime)\int\frac{d\theta d\theta^\prime}{(2\pi)^2}\sum_a\pi\alpha^2_{a;k_\mathrm{F},\theta,\theta^\prime}D_a\left(2k_\mathrm{F}\sin\frac{\theta-\theta^\prime}{2},\omega-\omega^\prime\right).
\label{eq:sigmaME2}
\end{equation}
\end{widetext}

\begin{figure}[!t]
\vspace{2mm}
\centering
\includegraphics[width=0.92\columnwidth]{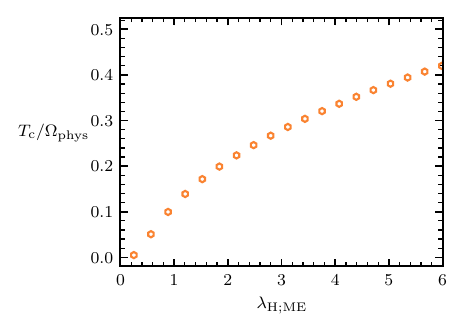}\hspace{3mm}
\caption{Transition temperature $T_\mathrm{c}$ in Migdal-Eliashberg theory calculations of the Holstein model divided by the physical (renormalized) phonon frequency $\Omega_\mathrm{phys}$ calculated as a function of the conventional Migdal-Eliashberg coupling constant $\lambda_\mathrm{H;ME}$ from Eq.~\eqref{eq:gapequationH}.}
\label{fig:tcHolstein}
\end{figure}

In the H model the phonon propagator and coupling coefficient are independent of momentum and its angle on the Fermi surface, and it is convenient to define the coupling constant as $\lambda_\mathrm{H;ME}=\lambda^0_\mathrm{H;ME}\frac{\Omega_0}{\Omega_\mathrm{phys}}$ using the physical, renormalized phonon frequency. After linearizing in the anomalous self energy we have, explicitly
\begin{widetext}
\begin{eqnarray}
(1-Z(\omega_\mathrm{n}))\omega_\mathrm{n}&=&-\frac{\pi\lambda_\mathrm{H;ME}}{2}~T\sum_{\Omega_\mathrm{n}}\sgn(\Omega_\mathrm{n})\frac{2\Omega_\mathrm{phys}^2}{(\omega_\mathrm{n}-\Omega_\mathrm{n})^2+\Omega_\mathrm{phys}^2},
\label{eq:massH}
\\
W(\omega_\mathrm{n})&=&\frac{\pi\lambda_\mathrm{H;ME}}{2}~T\sum_{\Omega_\mathrm{n}}\frac{W(\Omega_\mathrm{n})}{\left|\Omega_\mathrm{n} Z(\Omega_\mathrm{n})\right|}\frac{2\Omega_\mathrm{phys}^2}{(\omega_\mathrm{n}-\Omega_\mathrm{n})^2+\Omega_\mathrm{phys}^2}.
\label{eq:gapequationH}
\end{eqnarray}
\end{widetext}
These are the forms in which the Migdal-Eliashberg equations for the electron self energy are traditionally presented and solved. If the temperature is low compared to the phonon frequency one finds from Eq.~\eqref{eq:massH} that the low frequency mass enhancement is $1+\lambda_\mathrm{H;ME}$. The critical transition temperature may easily be determined following Bergmann and Rainer~\cite{Bergmann73} by recasting Eq. ~\eqref{eq:gapequationH} as an eigenvalue equation for the vector $W(\omega_\mathrm{n})/|\omega_\mathrm{n} Z(\omega_\mathrm{n})|$ and defining the transition temperature $T_\mathrm{c}$ as the temperature at which the leading eigenvalue vanishes.  The result is a $T_\mathrm{c}$ which, {\em as a fraction of the renormalized phonon frequency}, evolves from the small $\lambda$ Bardeen-Cooper-Schrieffer (BCS) form of $e^{-1/\lambda}$ to the Allen-Dynes  form of $\sqrt{\lambda}$ as $\lambda$ is increased from small to large values. Figure~\ref{fig:tcHolstein} shows the ratio of $T_\mathrm{c}$ to $\Omega_\mathrm{phys}$ as a function of $\lambda_\mathrm{H;ME}$ calculated from Eq.~\eqref{eq:gapequationH}. In the present context it is of greater relevance to present the results as the ratio of $T_\mathrm{c}$ to the bare oscillator frequency $\Omega_0$, as  a function of the bare coupling  $\lambda^0_\mathrm{H;ME}$. These are obtained by a simple scaling of the results in Fig.~\ref{fig:tcHolstein} and are presented in the main text in Fig.~\ref{fig:Fig1}.

\begin{figure*}
\vspace{1mm}
\centering
\includegraphics[width=2\columnwidth]{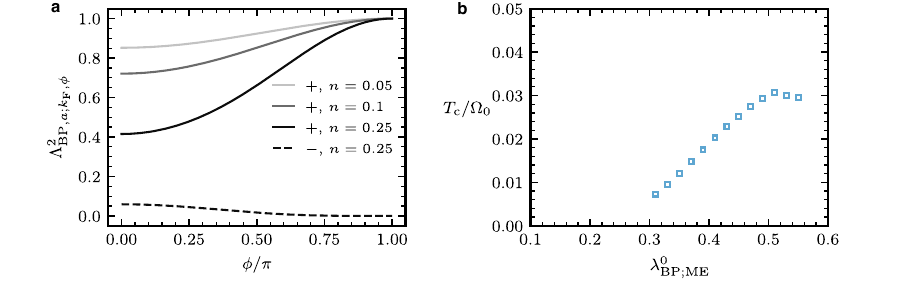} 
\caption{{\bf a.} Momentum dependence of the electron-phonon coupling parameter in Eq.~\eqref{alphaBPdef} for the $+$ mode of the bond-Peierls model calculated for densities $n=0.05$, $0.1$, and $0.25$ carrier per site  (solid lines; highest to lowest) and for the $-$ mode of the model for density $n=0.25$ carrier per site (dashed line). {\bf b.} Transition temperature $T_\mathrm{c}$ in Migdal-Eliashberg theory calculations of the bond-Peierls model divided by bare phonon frequency $\Omega_0$ as function of the bare coupling $\lambda_\mathrm{BP;ME}^0$ for density $n=0.25$ carrier per site.}
\label{fig:coupling}
\vspace{-3mm}
\end{figure*}

In the BP model the presence of two phonon modes in the unit cell and the momentum dependence of their coupling coefficients  and phonon frequencies make the analysis more involved. In order to evaluate Eq.~\eqref{eq:sigmaME1} for the BP model we need the coupling function $\Lambda(k_x,k_y)$ (Eq.~\eqref{eq:Lambdapmdef}) at the momentum $(k+k^\prime)/2$. For simplicity, we make the circular Fermi surface approximation and define $\psi=(\theta+\theta^\prime)/2$ and $\phi=\theta-\theta^\prime$, then Eq.~\eqref{eq:sigmaME2} can be written as
\begin{widetext}
\begin{equation}
\Sigma(\omega)=\frac{\pi \lambda^0_\mathrm{BP}}{2}\sum_{a=\pm}T\sum_{\omega^\prime}\mathbf{\mathcal{G}}(\omega^\prime)\int\frac{d\phi}{2\pi}\Lambda^2_{\mathrm{BP},a;k_\mathrm{F},\phi}2D_a\left(2k_\mathrm{F}\sin\frac{\phi}{2},\omega-\omega^\prime\right),
\label{eq:sigmaMEBP}
\end{equation}
with
\begin{equation}
\Lambda^2_{\mathrm{BP},\pm;k_\mathrm{F},\phi}=\frac{1}{8}\int\frac{d\psi}{2\pi}\Bigg(\cos (k_\mathrm{F}\cos\psi \cos\frac{\phi}{2}) \pm \cos(k_\mathrm{F}\sin\psi \cos\frac{\phi}{2}) \Bigg)^2.
\label{alphaBPdef}
\end{equation}
\end{widetext}
Figure~\ref{fig:coupling}{\bf a} shows the dependence of the coupling constant for the $+$ mode as a function of phonon momentum (parametrized by $\phi$) calculated within the circular Fermi surface approximation for densities $n = 0.25$, $0.1$ and $0.05$. Also shown is the coupling for the $-$ mode at $n=0.25$ (for the lower densities the coupling function is essentially indistinguishable from zero). We see that at all of the relevant densities  the $-$ mode decouples.  We solve the linearized gap equations for the BP model in the explicit form 
\begin{widetext}
\begin{eqnarray}
(1-Z(\omega_\mathrm{n}))\omega_\mathrm{n}&=&-~\frac{\pi \lambda_\mathrm{BP;ME}^0}{2}T\sum_{\Omega_n}\sgn(\Omega_\mathrm{n})2\mathcal{D}(\omega_\mathrm{n}-\Omega_\mathrm{n}),
\label{ZBPfinal}\\
W(\omega_\mathrm{n})&=&\frac{\pi \lambda_\mathrm{BP;ME}^0}{2}~ T\sum_{\Omega_\mathrm{m}}\frac{W(\Omega_\mathrm{m})}{\left|\Omega_\mathrm{m}Z(\Omega_\mathrm{m})\right|}2\mathcal{D}(\omega_\mathrm{n}-\Omega_\mathrm{m}),
\label{WBPfinal}
\end{eqnarray}
with
\begin{equation}
\mathcal{D}(\Omega_\mathrm{n} - \Omega_\mathrm{m})=\int\frac{d\phi}{2\pi}\Lambda^2_{\mathrm{BP},\pm;k_\mathrm{F},\phi}\frac{\Omega_0\Omega_+(2k_\mathrm{F} \sin\frac{\phi}{2})}{(\omega_n-\Omega_m)^2+\Omega_+^2(2k_\mathrm{F} \sin\frac{\phi}{2})}.
\end{equation}
\end{widetext}
We calculate the transition temperature by  proceeding as in the H case. Note however that we have formulated the equations form the outset in terms of the bare $\lambda$, $\lambda^0$, not the conventional ME $\lambda$ defined in terms of the renormalized phonon frequency $\Omega_\mathrm{phys}$. Results for $n=0.25$ are shown in the right panel of Fig.~\ref{fig:coupling}{\bf b} and in the main text in  Fig.~\ref{fig:Fig1}.

\section{McMillan's phenomenological approach to phonon-mediated strong-coupling superconductivity} \label{Appendix:AppE}

McMillan's approach to strong-coupling superconductivity is based on a phenomenological treatment of experimental data on conventional superconducting materials within the framework of Migdal-Eliashberg theory. This approach makes use of a coupling constant $\lambda$ estimated directly from experiment by considering an electron-phonon interaction averaged over the Fermi surface, and supplements this treatment by empirical parameters used to mimic the effect of Coulomb interaction in order to better fit experimental data. The Migdal-Eliashberg theory itself is valid in the adiabatic limit $t \gg \Omega $ and for moderate values of $\lambda$, because at $\lambda  \gtrsim 1$, apart from structural instability~\cite{CohenBounds}, the Fermi liquid becomes a metastable state, higher in energy than a state formed of bipolarons~\cite{ChakravertyBipolaron,ME_CDW1,ME_CDW2,MEbreakdownAlexandrov,MEBreakdownMillis,MEBreakdownKivelson,TcBoundKivelson,GeneralHolsteinBipolaron}. Within its domain of applicability,  McMillan's formula finds that strong electron-phonon coupling induces a superconducting instability out of a Fermi liquid at~\cite{McMillanFormula}
\begin{equation}
\frac{T_\mathrm{c}}{\Omega} = \frac{1}{1.45} e^{  -1.04 \frac{1+\lambda}{\lambda-\mu^\star(1 + 0.62\lambda)}},
\label{Eq:McMillan}
\end{equation}
where $\mu^\star=0.12$ is the value of the Coulomb pseudopotential found in many materials~\cite{McMillanFormula}. A typical upper bound from McMillan's approach can thus be estimated at about $\lambda = 1$ to give a maximum $T_\mathrm{c}/\Omega \sim 0.05$, in qualitative agreement with the results of  Migdal-Eliashberg theory applied to the two models presented in  Fig.~\ref{fig:Fig1} (see also Appendix~\ref{Appendix:AppD}). These comparisons illustrate that $T_{\mathrm{c}}$ obtained in our model of bipolaronic superconductivity generically exceeds previously held expectations.

\section{Effective electronic Hamiltonian in the antiadiabatic limit}\label{Appendix:AppF}

To unravel the pairing mechanism responsible for the formation of bipolarons, we derive an effective electronic Hamiltonian in the antiadiabatic limit $t \ll \Omega$ by projecting out high-energy subspaces with one or more phonons~\cite{SousBipolaron,Takahashi_1977}. This procedure is valid at strong coupling $\lambda \gg 1$ within the antiadiabatic regime if $t \ll \alpha \ll \Omega$ such that $\alpha^2 \gg \Omega t$. To second order, we find an effective two-electron Hamiltonian given by:
\begin{equation}
\hat{\cal H}_\mathrm{eff.} = \hat{h}_\mathrm{e} + \hat{\cal U}_{\mathrm{e\mbox{-}e}} + \hat{\cal V}_{\mathrm{e\mbox{-}e}},
\end{equation}
where
\begin{eqnarray}
\hat{h}_\mathrm{e} &=& -\epsilon \sum_{i,\sigma}^{}\hat{n}_{i,\sigma} - t
\sum_{\langle i,j \rangle,\sigma}^{} \left( \hat{c}_{i,\sigma}^\dagger \hat{c}_{j,\sigma} +  \mathrm{h.c.}\right), \\
\hat{\cal U}_{\mathrm{e\mbox{-}e}}  &=&  \tilde{U} \sum_{i} \hat{n}_{i,\uparrow} \hat{n}_{i,\downarrow} -T\sum_{i}^{}\left[
\hat{c}^\dagger_{i,\uparrow} \hat{c}^\dagger_{i,\downarrow} \hat{c}_{j,\downarrow} \hat{c}_{j,\uparrow}+ \mathrm{h.c.}\right] ,\nonumber \\ \\
\hat{\cal V}_{\mathrm{e\mbox{-}e}} &=&   \tilde{V} \sum_{\langle i,j \rangle} \hat{n}_{i} \hat{n}_{j}  + J \sum_{\langle i,j \rangle} \vec{S}_i \cdot \vec{S}_j,
\end{eqnarray}
where $\hat{n}_i = \hat{n}_{i,\uparrow} + \hat{n}_{i,\downarrow}$, $\vec{S}_i = \frac{1}{2} \sum_{\alpha,\beta} \hat{c}_{i,\alpha}^\dagger \vec{\sigma}_{\alpha,\beta}\hat{c}_{i,\beta}$ and $\vec{\sigma}$ is a vector of $\mathrm{SU(2)}$ Pauli operators. Here $\epsilon = \frac{4\alpha^2}{\Omega}$ is the polaron formation energy, $\tilde{U} = U - \frac{8\alpha^2}{\Omega -U} + \frac{8\alpha^2}{\Omega}$ is the amplitude of a phonon-renormalized effective onsite density-density interaction, $T = \frac{2\alpha^2}{\Omega-U}$ is the amplitude of a phonon-mediated onsite electron pair hopping interaction~\cite{SousBipolaron}, $\tilde{V} =\frac{2\alpha^2}{\Omega} - \frac{\alpha^2}{\Omega + U}$ is the amplitude of a phonon-mediated nearest-neighbor density-density interaction~\cite{SousScRep}, and $J = \frac{4\alpha^2}{\Omega + U}$ is the amplitude of a phonon-mediated $\mathrm{SU(2)}$-preserving   nearest-neighbor spin-spin interaction~\cite{SousScRep,SousBipolaron}.

Our numerical results away from the antiadiabatic limit indicate that the salient features embodied by $\hat{\cal H}_\mathrm{eff.}$, specifically the phonon-mediated  kinetic energy enhancing pair-hopping interaction~\cite{SousBipolaron},   continue to hold qualitatively as $t/\Omega$ increases, as evidenced by the light bipolaron masses.   In contrast, in the  adiabatic $t / \Omega \gg 1$, phonons should behave classically and have no dynamics, thus a pair of electrons experience a retarded phonon-mediated attraction and  form a singlet bipolaron localized on a lattice bond in order to minimize the total energy. Our numerical results indicate that away from these asymptotic limits a competition between the phonon-mediated kinetic energy enhancing electron pair-hopping  interaction and the tendency to localize electron pairs is at play and determines the fate of bipolarons, which, nonetheless, appear to be light in a large region of parameter space.

Finally, we see that, at least in the antiadiabatic limit, the $\hat{\cal V}_{\mathrm{e\mbox{-}e}}$ contains a phonon-mediated repulsive part~\cite{SousScRep}, which disfavors pairing  in all but the rotationally symmetric $s$-channel, and 
a large $U/t$, e.g. $U=8t$ discussed in the main text, further enhances this tendency.

\section{Bond-Peierls coupling in the iron-based pnictide superconductors}\label{Appendix:AppG}

The bond-Peierls electron-phonon coupling arises generically in systems where the orbitals of out-of-plane atoms mix with the bonding orbitals of in-plane atoms~\cite{Kagan}.  Here, transverse fluctuations of the displacement of the out-of-plane atoms give rise to modulation of the barrier for electron tunneling across bonds, precisely as embodied by the model in Eqs.~\eqref{Eq:H}, \eqref{Eq:Veph}.  Interestingly, Refs.~\cite{HauleNatMater,PnictogenHeight} show that this physics is operative in the iron pnictides wherein  modulation of the pnictogen height can strongly modulate particular hopping pathways. Figure~\ref{fig:Fig5} demonstrates this scenario. Here a pnictogen atom sits at the apex of an octahedron with four iron atoms residing in the $x$-$y$ plane in the middle of the octahedron. The transverse motion of the pnictogen atom out of the $x$-$y$ plane in the $z$ direction causes fluctuations in the barrier for electronic tunneling between the iron atoms within the $x$-$y$ plane.~\cite{HauleNatMater,PnictogenHeight} (Fig.~\ref{fig:Fig5}{\bf a}). A dominant pathway for direct electronic hopping  between iron atoms in this class of materials involves overlaps between lopes of $d_{xy}$ orbitals of opposite sign on neighboring iron atoms in the $x$-$y$ plane, resulting in a negative hopping $t<0$~\cite{HauleNatMater}. An indirect electronic pathway, resulting from a second order, superexchange-like process, involves the overlap of the apex atom's $p_x$ orbital with each lobe of the two $d_{xy}$ orbitals, resulting in a net positive hopping $t'>0$~\cite{HauleNatMater}. These two pathways  (Fig.~\ref{fig:Fig5}{\bf b}) nearly cancel in FeSe because the magnitudes of $t$ and $t'$ are nearly equal, and more generally the ratio of $t$ to $t'$ vary in other pnictide materials resulting in an overall reduction in the magnitude of the net electronic hopping between the iron atoms~\cite{HauleNatMater}. This interference effect combined with the large modulation of the tunneling barrier by the displacement of the pnictogen atom along this particular hopping pathway means that the value of the dimensionless electron-bond-phonon coupling strength $\lambda$ relevant to this mode in this class of materials can be large, see Refs.~\cite{ZXSHENFESE,HauleNatMater}. As an example, Ref.~\cite{ZXSHENFESE0} suggests a value of $\lambda \sim 0.5$ in one member of this family of materials, but more work is needed to accurately determine the strength of electron-phonon coupling in specific compounds. The net overall hopping in the $x$-$y$ plane is roughly $\sim 50$~meV~\cite{HauleNatMater} and the transverse phonon frequency is estimated to be $\sim 5.3$ THz $\approx 22$ meV in FeSe, and $\sim 17$ meV in FeTe~\cite{ZXSHENFESE}, implying a ratio of the relevant phonon frequency to the magnitude of the relevant (net) electron hopping of $\sim 2$ - $3$ in these materials.
We also note that while the pairing symmetry in these materials has not yet been fully determined, an ``extended $s$-wave'' state with some similarities to our bond bipolaron state is a leading candidate.

This analysis reveals that the bond-Peierls coupling may be operative in the pnictides.  However, additional electron-phonon interaction terms may exist in these materials. For example, since a bond between two iron atoms connects two octahedra, the motion  of a single pnictogen atom out of plane within one octahedron is correlated with that of another pnictogen atom in the  neighboring octahedron, and this correlated pnictogen-pnictogen motion can simultaneously modulate the hopping across two iron-iron bonds, giving rise to yet another electron-phonon coupling term in these materials. The true extent to which the bond-Peierls coupling is important in determining the behavior of these materials requires more work to understand the interplay between the electron-phonon interaction terms and other features such as those presented by the multiple electronic bands, Hund’s coupling, and the form and range of the effective electron-electron interactions near the Fermi surface.

\catcode`'=9
\catcode``=9

\begin{thebibliography}{60}%
\makeatletter
\providecommand \@ifxundefined [1]{%
 \@ifx{#1\undefined}
}%
\providecommand \@ifnum [1]{%
 \ifnum #1\expandafter \@firstoftwo
 \else \expandafter \@secondoftwo
 \fi
}%
\providecommand \@ifx [1]{%
 \ifx #1\expandafter \@firstoftwo
 \else \expandafter \@secondoftwo
 \fi
}%
\providecommand \natexlab [1]{#1}%
\providecommand \enquote  [1]{``#1''}%
\providecommand \bibnamefont  [1]{#1}%
\providecommand \bibfnamefont [1]{#1}%
\providecommand \citenamefont [1]{#1}%
\providecommand \href@noop [0]{\@secondoftwo}%
\providecommand \href [0]{\begingroup \@sanitize@url \@href}%
\providecommand \@href[1]{\@@startlink{#1}\@@href}%
\providecommand \@@href[1]{\endgroup#1\@@endlink}%
\providecommand \@sanitize@url [0]{\catcode `\\12\catcode `\$12\catcode
  `\&12\catcode `\#12\catcode `\^12\catcode `\_12\catcode `\%12\relax}%
\providecommand \@@startlink[1]{}%
\providecommand \@@endlink[0]{}%
\providecommand \url  [0]{\begingroup\@sanitize@url \@url }%
\providecommand \@url [1]{\endgroup\@href {#1}{\urlprefix }}%
\providecommand \urlprefix  [0]{URL }%
\providecommand \Eprint [0]{\href }%
\providecommand \doibase [0]{http://dx.doi.org/}%
\providecommand \selectlanguage [0]{\@gobble}%
\providecommand \bibinfo  [0]{\@secondoftwo}%
\providecommand \bibfield  [0]{\@secondoftwo}%
\providecommand \translation [1]{[#1]}%
\providecommand \BibitemOpen [0]{}%
\providecommand \bibitemStop [0]{}%
\providecommand \bibitemNoStop [0]{.\EOS\space}%
\providecommand \EOS [0]{\spacefactor3000\relax}%
\providecommand \BibitemShut  [1]{\csname bibitem#1\endcsname}%
\let\auto@bib@innerbib\@empty
\bibitem [{\citenamefont {Ashcroft}(1968)}]{HighPressure1}%
  \BibitemOpen
  \bibfield  {author} {\bibinfo {author} {\bibfnamefont {N.~W.}\ \bibnamefont
  {Ashcroft}},\ }\bibfield  {title} {\enquote {\bibinfo {title} {Metallic
  hydrogen: {A} high-temperature superconductor?}}\ }\href {\doibase
  10.1103/PhysRevLett.21.1748} {\bibfield  {journal} {\bibinfo  {journal}
  {Phys. Rev. Lett.}\ }\textbf {\bibinfo {volume} {21}},\ \bibinfo {pages}
  {1748} (\bibinfo {year} {1968})}\BibitemShut {NoStop}%
\bibitem [{\citenamefont {Drozdov}\ \emph {et~al.}(2015)\citenamefont
  {Drozdov}, \citenamefont {Eremets}, \citenamefont {Troyan}, \citenamefont
  {Ksenofontov},\ and\ \citenamefont {Shylin}}]{HighPressure2}%
  \BibitemOpen
  \bibfield  {author} {\bibinfo {author} {\bibfnamefont {A.~P.}\ \bibnamefont
  {Drozdov}}, \bibinfo {author} {\bibfnamefont {M.~I.}\ \bibnamefont
  {Eremets}}, \bibinfo {author} {\bibfnamefont {I.~A.}\ \bibnamefont {Troyan}},
  \bibinfo {author} {\bibfnamefont {V.}~\bibnamefont {Ksenofontov}}, \ and\
  \bibinfo {author} {\bibfnamefont {S.~I.}\ \bibnamefont {Shylin}},\ }\bibfield
   {title} {\enquote {\bibinfo {title} {Conventional superconductivity at 203
  {K}elvin at high pressures in the sulfur hydride system},}\ }\href {\doibase
  10.1038/nature14964} {\bibfield  {journal} {\bibinfo  {journal} {Nature}\
  }\textbf {\bibinfo {volume} {525}},\ \bibinfo {pages} {73} (\bibinfo {year}
  {2015})}\BibitemShut {NoStop}%
\bibitem [{\citenamefont {Zhou}\ \emph {et~al.}(2021)\citenamefont {Zhou},
  \citenamefont {Lee}, \citenamefont {Imada}, \citenamefont {Trivedi},
  \citenamefont {Phillips}, \citenamefont {Kee}, \citenamefont
  {T{\"o}rm{\"a}},\ and\ \citenamefont {Eremets}}]{HighTcReview}%
  \BibitemOpen
  \bibfield  {author} {\bibinfo {author} {\bibfnamefont {X.}~\bibnamefont
  {Zhou}}, \bibinfo {author} {\bibfnamefont {W.-S.}\ \bibnamefont {Lee}},
  \bibinfo {author} {\bibfnamefont {M.}~\bibnamefont {Imada}}, \bibinfo
  {author} {\bibfnamefont {N.}~\bibnamefont {Trivedi}}, \bibinfo {author}
  {\bibfnamefont {P.}~\bibnamefont {Phillips}}, \bibinfo {author}
  {\bibfnamefont {H.-Y.}\ \bibnamefont {Kee}}, \bibinfo {author} {\bibfnamefont
  {P.}~\bibnamefont {T{\"o}rm{\"a}}}, \ and\ \bibinfo {author} {\bibfnamefont
  {M.}~\bibnamefont {Eremets}},\ }\bibfield  {title} {\enquote {\bibinfo
  {title} {High-temperature superconductivity},}\ }\href {\doibase
  10.1038/s42254-021-00324-3} {\bibfield  {journal} {\bibinfo  {journal} {Nat.
  Rev. Phys.}\ }\textbf {\bibinfo {volume} {1}} (\bibinfo {year} {2021}),\
  10.1038/s42254-021-00324-3}\BibitemShut {NoStop}%
\bibitem [{\citenamefont {Chakraverty}(1979)}]{ChakravertyBipolaron}%
  \BibitemOpen
  \bibfield  {author} {\bibinfo {author} {\bibfnamefont {B.~K.}\ \bibnamefont
  {Chakraverty}},\ }\bibfield  {title} {\enquote {\bibinfo {title} {Possibility
  of insulator to superconductor phase transition},}\ }\href {\doibase
  10.1051/jphyslet:0197900400509900} {\bibfield  {journal} {\bibinfo  {journal}
  {J. Phys. (Paris) Lett.}\ }\textbf {\bibinfo {volume} {40}},\ \bibinfo
  {pages} {99} (\bibinfo {year} {1979})}\BibitemShut {NoStop}%
\bibitem [{\citenamefont {Scalettar}\ \emph {et~al.}(1989)\citenamefont
  {Scalettar}, \citenamefont {Bickers},\ and\ \citenamefont
  {Scalapino}}]{ME_CDW1}%
  \BibitemOpen
  \bibfield  {author} {\bibinfo {author} {\bibfnamefont {R.~T.}\ \bibnamefont
  {Scalettar}}, \bibinfo {author} {\bibfnamefont {N.~E.}\ \bibnamefont
  {Bickers}}, \ and\ \bibinfo {author} {\bibfnamefont {D.~J.}\ \bibnamefont
  {Scalapino}},\ }\bibfield  {title} {\enquote {\bibinfo {title} {Competition
  of pairing and {P}eierls--charge-density-wave correlations in a
  two-dimensional electron-phonon model},}\ }\href {\doibase
  10.1103/PhysRevB.40.197} {\bibfield  {journal} {\bibinfo  {journal} {Phys.
  Rev. B}\ }\textbf {\bibinfo {volume} {40}},\ \bibinfo {pages} {197} (\bibinfo
  {year} {1989})}\BibitemShut {NoStop}%
\bibitem [{\citenamefont {Marsiglio}(1990)}]{ME_CDW2}%
  \BibitemOpen
  \bibfield  {author} {\bibinfo {author} {\bibfnamefont {F.}~\bibnamefont
  {Marsiglio}},\ }\bibfield  {title} {\enquote {\bibinfo {title} {Pairing and
  charge-density-wave correlations in the {H}olstein model at half-filling},}\
  }\href {\doibase 10.1103/PhysRevB.42.2416} {\bibfield  {journal} {\bibinfo
  {journal} {Phys. Rev. B}\ }\textbf {\bibinfo {volume} {42}},\ \bibinfo
  {pages} {2416} (\bibinfo {year} {1990})}\BibitemShut {NoStop}%
\bibitem [{\citenamefont {Moussa}\ and\ \citenamefont
  {Cohen}(2006)}]{CohenBounds}%
  \BibitemOpen
  \bibfield  {author} {\bibinfo {author} {\bibfnamefont {J.~E.}\ \bibnamefont
  {Moussa}}\ and\ \bibinfo {author} {\bibfnamefont {M.~L.}\ \bibnamefont
  {Cohen}},\ }\bibfield  {title} {\enquote {\bibinfo {title} {Two bounds on the
  maximum phonon-mediated superconducting transition temperature},}\ }\href
  {\doibase 10.1103/PhysRevB.74.094520} {\bibfield  {journal} {\bibinfo
  {journal} {Phys. Rev. B}\ }\textbf {\bibinfo {volume} {74}},\ \bibinfo
  {pages} {094520} (\bibinfo {year} {2006})}\BibitemShut {NoStop}%
\bibitem [{\citenamefont {Alexandrov}(2001)}]{MEbreakdownAlexandrov}%
  \BibitemOpen
  \bibfield  {author} {\bibinfo {author} {\bibfnamefont {A.~S.}\ \bibnamefont
  {Alexandrov}},\ }\bibfield  {title} {\enquote {\bibinfo {title} {Breakdown of
  the {M}igdal-{E}liashberg theory in the strong-coupling adiabatic regime},}\
  }\href {\doibase 10.1209/epl/i2001-00492-x} {\bibfield  {journal} {\bibinfo
  {journal} {Europhys. Lett.}\ }\textbf {\bibinfo {volume} {56}},\ \bibinfo
  {pages} {92} (\bibinfo {year} {2001})}\BibitemShut {NoStop}%
\bibitem [{\citenamefont {Werner}\ and\ \citenamefont
  {Millis}(2007)}]{MEBreakdownMillis}%
  \BibitemOpen
  \bibfield  {author} {\bibinfo {author} {\bibfnamefont {P.}~\bibnamefont
  {Werner}}\ and\ \bibinfo {author} {\bibfnamefont {A.~J.}\ \bibnamefont
  {Millis}},\ }\bibfield  {title} {\enquote {\bibinfo {title} {Efficient
  dynamical mean field simulation of the {H}olstein-{H}ubbard model},}\ }\href
  {\doibase 10.1103/PhysRevLett.99.146404} {\bibfield  {journal} {\bibinfo
  {journal} {Phys. Rev. Lett.}\ }\textbf {\bibinfo {volume} {99}},\ \bibinfo
  {pages} {146404} (\bibinfo {year} {2007})}\BibitemShut {NoStop}%
\bibitem [{\citenamefont {Esterlis}\ \emph
  {et~al.}(2018{\natexlab{a}})\citenamefont {Esterlis}, \citenamefont
  {Nosarzewski}, \citenamefont {Huang}, \citenamefont {Moritz}, \citenamefont
  {Devereaux}, \citenamefont {Scalapino},\ and\ \citenamefont
  {Kivelson}}]{MEBreakdownKivelson}%
  \BibitemOpen
  \bibfield  {author} {\bibinfo {author} {\bibfnamefont {I.}~\bibnamefont
  {Esterlis}}, \bibinfo {author} {\bibfnamefont {B.}~\bibnamefont
  {Nosarzewski}}, \bibinfo {author} {\bibfnamefont {E.~W.}\ \bibnamefont
  {Huang}}, \bibinfo {author} {\bibfnamefont {B.}~\bibnamefont {Moritz}},
  \bibinfo {author} {\bibfnamefont {T.~P.}\ \bibnamefont {Devereaux}}, \bibinfo
  {author} {\bibfnamefont {D.~J.}\ \bibnamefont {Scalapino}}, \ and\ \bibinfo
  {author} {\bibfnamefont {S.~A.}\ \bibnamefont {Kivelson}},\ }\bibfield
  {title} {\enquote {\bibinfo {title} {Breakdown of the {M}igdal-{E}liashberg
  theory: A determinant quantum {M}onte {C}arlo study},}\ }\href {\doibase
  10.1103/PhysRevB.97.140501} {\bibfield  {journal} {\bibinfo  {journal} {Phys.
  Rev. B}\ }\textbf {\bibinfo {volume} {97}},\ \bibinfo {pages} {140501}
  (\bibinfo {year} {2018}{\natexlab{a}})}\BibitemShut {NoStop}%
\bibitem [{\citenamefont {Chakraverty}\ \emph {et~al.}(1998)\citenamefont
  {Chakraverty}, \citenamefont {Ranninger},\ and\ \citenamefont
  {Feinberg}}]{CRFBipolaron}%
  \BibitemOpen
  \bibfield  {author} {\bibinfo {author} {\bibfnamefont {B.~K.}\ \bibnamefont
  {Chakraverty}}, \bibinfo {author} {\bibfnamefont {J.}~\bibnamefont
  {Ranninger}}, \ and\ \bibinfo {author} {\bibfnamefont {D.}~\bibnamefont
  {Feinberg}},\ }\bibfield  {title} {\enquote {\bibinfo {title} {Experimental
  and theoretical constraints of bipolaronic superconductivity in high
  ${T}_\mathrm{c}$ materials: An impossibility},}\ }\href {\doibase
  10.1103/PhysRevLett.81.433} {\bibfield  {journal} {\bibinfo  {journal} {Phys.
  Rev. Lett.}\ }\textbf {\bibinfo {volume} {81}},\ \bibinfo {pages} {433}
  (\bibinfo {year} {1998})}\BibitemShut {NoStop}%
\bibitem [{\citenamefont {Bon{\v{c}}a}\ \emph {et~al.}(2000)\citenamefont
  {Bon{\v{c}}a}, \citenamefont {Katras{\v{n}}ik},\ and\ \citenamefont
  {Trugman}}]{BoncaHBipolaron}%
  \BibitemOpen
  \bibfield  {author} {\bibinfo {author} {\bibfnamefont {J.}~\bibnamefont
  {Bon{\v{c}}a}}, \bibinfo {author} {\bibfnamefont {T.}~\bibnamefont
  {Katras{\v{n}}ik}}, \ and\ \bibinfo {author} {\bibfnamefont {S.~A.}\
  \bibnamefont {Trugman}},\ }\bibfield  {title} {\enquote {\bibinfo {title}
  {Mobile bipolaron},}\ }\href {\doibase 10.1103/PhysRevLett.84.315} {\bibfield
   {journal} {\bibinfo  {journal} {Phys. Rev. Lett.}\ }\textbf {\bibinfo
  {volume} {84}},\ \bibinfo {pages} {3153} (\bibinfo {year}
  {2000})}\BibitemShut {NoStop}%
\bibitem [{\citenamefont {Macridin}\ \emph {et~al.}(2004)\citenamefont
  {Macridin}, \citenamefont {Sawatzky},\ and\ \citenamefont
  {Jarrell}}]{MacridinBipolaron}%
  \BibitemOpen
  \bibfield  {author} {\bibinfo {author} {\bibfnamefont {A.}~\bibnamefont
  {Macridin}}, \bibinfo {author} {\bibfnamefont {G.~A.}\ \bibnamefont
  {Sawatzky}}, \ and\ \bibinfo {author} {\bibfnamefont {M.}~\bibnamefont
  {Jarrell}},\ }\bibfield  {title} {\enquote {\bibinfo {title} {Two-dimensional
  {H}ubbard-{H}olstein bipolaron},}\ }\href {\doibase
  10.1103/PhysRevB.69.245111} {\bibfield  {journal} {\bibinfo  {journal} {Phys.
  Rev. B}\ }\textbf {\bibinfo {volume} {69}},\ \bibinfo {pages} {245111}
  (\bibinfo {year} {2004})}\BibitemShut {NoStop}%
\bibitem [{\citenamefont {Alexandrov}\ and\ \citenamefont
  {Kornilovitch}(1999)}]{PachaPolaron}%
  \BibitemOpen
  \bibfield  {author} {\bibinfo {author} {\bibfnamefont {A.~S.}\ \bibnamefont
  {Alexandrov}}\ and\ \bibinfo {author} {\bibfnamefont {P.~E.}\ \bibnamefont
  {Kornilovitch}},\ }\bibfield  {title} {\enquote {\bibinfo {title} {Mobile
  small polaron},}\ }\href {\doibase 10.1103/PhysRevLett.82.807} {\bibfield
  {journal} {\bibinfo  {journal} {Phys. Rev. Lett.}\ }\textbf {\bibinfo
  {volume} {82}},\ \bibinfo {pages} {807} (\bibinfo {year} {1999})}\BibitemShut
  {NoStop}%
\bibitem [{\citenamefont {Bon\ifmmode~\check{c}\else \v{c}\fi{}a}\ and\
  \citenamefont {Trugman}(2001)}]{BoncaEH}%
  \BibitemOpen
  \bibfield  {author} {\bibinfo {author} {\bibfnamefont {J.}~\bibnamefont
  {Bon\ifmmode~\check{c}\else \v{c}\fi{}a}}\ and\ \bibinfo {author}
  {\bibfnamefont {S.~A.}\ \bibnamefont {Trugman}},\ }\bibfield  {title}
  {\enquote {\bibinfo {title} {Bipolarons in the extended {H}olstein {H}ubbard
  model},}\ }\href {\doibase 10.1103/PhysRevB.64.094507} {\bibfield  {journal}
  {\bibinfo  {journal} {Phys. Rev. B}\ }\textbf {\bibinfo {volume} {64}},\
  \bibinfo {pages} {094507} (\bibinfo {year} {2001})}\BibitemShut {NoStop}%
\bibitem [{\citenamefont {Hague}\ \emph {et~al.}(2007)\citenamefont {Hague},
  \citenamefont {Kornilovitch}, \citenamefont {Samson},\ and\ \citenamefont
  {Alexandrov}}]{Pacha1}%
  \BibitemOpen
  \bibfield  {author} {\bibinfo {author} {\bibfnamefont {J.~P.}\ \bibnamefont
  {Hague}}, \bibinfo {author} {\bibfnamefont {P.~E.}\ \bibnamefont
  {Kornilovitch}}, \bibinfo {author} {\bibfnamefont {J.~H.}\ \bibnamefont
  {Samson}}, \ and\ \bibinfo {author} {\bibfnamefont {A.~S.}\ \bibnamefont
  {Alexandrov}},\ }\bibfield  {title} {\enquote {\bibinfo {title} {Superlight
  small bipolarons in the presence of a strong {C}oulomb repulsion},}\ }\href
  {\doibase 10.1103/PhysRevLett.98.037002} {\bibfield  {journal} {\bibinfo
  {journal} {Phys. Rev. Lett.}\ }\textbf {\bibinfo {volume} {98}},\ \bibinfo
  {pages} {037002} (\bibinfo {year} {2007})}\BibitemShut {NoStop}%
\bibitem [{\citenamefont {Chandler}\ and\ \citenamefont
  {Marsiglio}(2014)}]{Chandler}%
  \BibitemOpen
  \bibfield  {author} {\bibinfo {author} {\bibfnamefont {C.~J.}\ \bibnamefont
  {Chandler}}\ and\ \bibinfo {author} {\bibfnamefont {F.}~\bibnamefont
  {Marsiglio}},\ }\bibfield  {title} {\enquote {\bibinfo {title} {Extended
  versus standard {H}olstein model: {R}esults in two and three dimensions},}\
  }\href {\doibase 10.1103/PhysRevB.90.125131} {\bibfield  {journal} {\bibinfo
  {journal} {Phys. Rev. B}\ }\textbf {\bibinfo {volume} {90}},\ \bibinfo
  {pages} {125131} (\bibinfo {year} {2014})}\BibitemShut {NoStop}%
\bibitem [{\citenamefont {Bari\ifmmode \check{s}\else
  \v{s}\fi{}i\ifmmode~\acute{c}\else \'{c}\fi{}}\ \emph
  {et~al.}(1970)\citenamefont {Bari\ifmmode \check{s}\else
  \v{s}\fi{}i\ifmmode~\acute{c}\else \'{c}\fi{}}, \citenamefont {Labb\'e},\
  and\ \citenamefont {Friedel}}]{BarisicPeierls1}%
  \BibitemOpen
  \bibfield  {author} {\bibinfo {author} {\bibfnamefont {S.}~\bibnamefont
  {Bari\ifmmode \check{s}\else \v{s}\fi{}i\ifmmode~\acute{c}\else \'{c}\fi{}}},
  \bibinfo {author} {\bibfnamefont {J.}~\bibnamefont {Labb\'e}}, \ and\
  \bibinfo {author} {\bibfnamefont {J.}~\bibnamefont {Friedel}},\ }\bibfield
  {title} {\enquote {\bibinfo {title} {Tight binding and transition-metal
  superconductivity},}\ }\href {\doibase 10.1103/PhysRevLett.25.919} {\bibfield
   {journal} {\bibinfo  {journal} {Phys. Rev. Lett.}\ }\textbf {\bibinfo
  {volume} {25}},\ \bibinfo {pages} {919} (\bibinfo {year} {1970})}\BibitemShut
  {NoStop}%
\bibitem [{\citenamefont {Su}\ \emph {et~al.}(1979)\citenamefont {Su},
  \citenamefont {Schrieffer},\ and\ \citenamefont {Heeger}}]{su1979solitons}%
  \BibitemOpen
  \bibfield  {author} {\bibinfo {author} {\bibfnamefont {W.~P.}\ \bibnamefont
  {Su}}, \bibinfo {author} {\bibfnamefont {J.~R.}\ \bibnamefont {Schrieffer}},
  \ and\ \bibinfo {author} {\bibfnamefont {A.~J.}\ \bibnamefont {Heeger}},\
  }\bibfield  {title} {\enquote {\bibinfo {title} {Solitons in
  polyacetylene},}\ }\href {\doibase 10.1103/PhysRevLett.42.1698} {\bibfield
  {journal} {\bibinfo  {journal} {Phys. Rev. Lett.}\ }\textbf {\bibinfo
  {volume} {42}},\ \bibinfo {pages} {1698} (\bibinfo {year}
  {1979})}\BibitemShut {NoStop}%
\bibitem [{\citenamefont {Capone}\ \emph {et~al.}(1997)\citenamefont {Capone},
  \citenamefont {Stephan},\ and\ \citenamefont {Grilli}}]{Capone1}%
  \BibitemOpen
  \bibfield  {author} {\bibinfo {author} {\bibfnamefont {M.}~\bibnamefont
  {Capone}}, \bibinfo {author} {\bibfnamefont {W.}~\bibnamefont {Stephan}}, \
  and\ \bibinfo {author} {\bibfnamefont {M.}~\bibnamefont {Grilli}},\
  }\bibfield  {title} {\enquote {\bibinfo {title} {Small-polaron formation and
  optical absorption in {S}u-{S}chrieffer-{H}eeger and {H}olstein models},}\
  }\href {\doibase 10.1103/PhysRevB.56.4484} {\bibfield  {journal} {\bibinfo
  {journal} {Phys. Rev. B}\ }\textbf {\bibinfo {volume} {56}},\ \bibinfo
  {pages} {4484} (\bibinfo {year} {1997})}\BibitemShut {NoStop}%
\bibitem [{\citenamefont {Perroni}\ \emph {et~al.}(2004)\citenamefont
  {Perroni}, \citenamefont {Piegari}, \citenamefont {Capone},\ and\
  \citenamefont {Cataudella}}]{Capone2}%
  \BibitemOpen
  \bibfield  {author} {\bibinfo {author} {\bibfnamefont {C.~A.}\ \bibnamefont
  {Perroni}}, \bibinfo {author} {\bibfnamefont {E.}~\bibnamefont {Piegari}},
  \bibinfo {author} {\bibfnamefont {M.}~\bibnamefont {Capone}}, \ and\ \bibinfo
  {author} {\bibfnamefont {V.}~\bibnamefont {Cataudella}},\ }\bibfield  {title}
  {\enquote {\bibinfo {title} {Polaron formation for nonlocal electron-phonon
  coupling: {A} variational wave-function study},}\ }\href {\doibase
  10.1103/PhysRevB.69.174301} {\bibfield  {journal} {\bibinfo  {journal} {Phys.
  Rev. B}\ }\textbf {\bibinfo {volume} {69}},\ \bibinfo {pages} {174301}
  (\bibinfo {year} {2004})}\BibitemShut {NoStop}%
\bibitem [{\citenamefont {Marchand}\ \emph {et~al.}(2010)\citenamefont
  {Marchand}, \citenamefont {De~Filippis}, \citenamefont {Cataudella},
  \citenamefont {Berciu}, \citenamefont {Nagaosa}, \citenamefont {Prokof’ev},
  \citenamefont {Mishchenko},\ and\ \citenamefont {Stamp}}]{marchand2010sharp}%
  \BibitemOpen
  \bibfield  {author} {\bibinfo {author} {\bibfnamefont {D.~J.~J.}\
  \bibnamefont {Marchand}}, \bibinfo {author} {\bibfnamefont {G.}~\bibnamefont
  {De~Filippis}}, \bibinfo {author} {\bibfnamefont {V.}~\bibnamefont
  {Cataudella}}, \bibinfo {author} {\bibfnamefont {M.}~\bibnamefont {Berciu}},
  \bibinfo {author} {\bibfnamefont {N.}~\bibnamefont {Nagaosa}}, \bibinfo
  {author} {\bibfnamefont {N.~V.}\ \bibnamefont {Prokof’ev}}, \bibinfo
  {author} {\bibfnamefont {A.~S.}\ \bibnamefont {Mishchenko}}, \ and\ \bibinfo
  {author} {\bibfnamefont {P.~C.~E.}\ \bibnamefont {Stamp}},\ }\bibfield
  {title} {\enquote {\bibinfo {title} {Sharp transition for single polarons in
  the one-dimensional {S}u-{S}chrieffer-{H}eeger model},}\ }\href
  {https://link.aps.org/doi/10.1103/PhysRevLett.105.266605} {\bibfield
  {journal} {\bibinfo  {journal} {Phys. Rev. Lett.}\ }\textbf {\bibinfo
  {volume} {105}},\ \bibinfo {pages} {266605} (\bibinfo {year}
  {2010})}\BibitemShut {NoStop}%
\bibitem [{\citenamefont {Zhang}\ \emph {et~al.}(2021)\citenamefont {Zhang},
  \citenamefont {Prokof’ev},\ and\ \citenamefont
  {Svistunov}}]{ChaoPeierlsPolaron}%
  \BibitemOpen
  \bibfield  {author} {\bibinfo {author} {\bibfnamefont {C.}~\bibnamefont
  {Zhang}}, \bibinfo {author} {\bibfnamefont {N.~V.}\ \bibnamefont
  {Prokof’ev}}, \ and\ \bibinfo {author} {\bibfnamefont {B.~V.}\ \bibnamefont
  {Svistunov}},\ }\bibfield  {title} {\enquote {\bibinfo {title}
  {Peierls/{S}u-{S}chrieffer-{H}eeger polarons in two dimensions},}\ }\href
  {\doibase 10.1103/PhysRevB.104.035143} {\bibfield  {journal} {\bibinfo
  {journal} {Phys. Rev. B}\ }\textbf {\bibinfo {volume} {104}},\ \bibinfo
  {pages} {035143} (\bibinfo {year} {2021})}\BibitemShut {NoStop}%
\bibitem [{\citenamefont {Carbone}\ \emph {et~al.}(2021)\citenamefont
  {Carbone}, \citenamefont {Millis}, \citenamefont {Reichman},\ and\
  \citenamefont {Sous}}]{CarbonePeierlsPolaron}%
  \BibitemOpen
  \bibfield  {author} {\bibinfo {author} {\bibfnamefont {M.~R.}\ \bibnamefont
  {Carbone}}, \bibinfo {author} {\bibfnamefont {A.~J.}\ \bibnamefont {Millis}},
  \bibinfo {author} {\bibfnamefont {D.~R.}\ \bibnamefont {Reichman}}, \ and\
  \bibinfo {author} {\bibfnamefont {J.}~\bibnamefont {Sous}},\ }\bibfield
  {title} {\enquote {\bibinfo {title} {Bond-{P}eierls polaron: {M}oderate mass
  enhancement and current-carrying ground state},}\ }\href {\doibase
  10.1103/PhysRevB.104.L140307} {\bibfield  {journal} {\bibinfo  {journal}
  {Phys. Rev. B}\ }\textbf {\bibinfo {volume} {104}},\ \bibinfo {pages}
  {L140307} (\bibinfo {year} {2021})}\BibitemShut {NoStop}%
\bibitem [{\citenamefont {Sous}\ \emph {et~al.}(2018)\citenamefont {Sous},
  \citenamefont {Chakraborty}, \citenamefont {Krems},\ and\ \citenamefont
  {Berciu}}]{SousBipolaron}%
  \BibitemOpen
  \bibfield  {author} {\bibinfo {author} {\bibfnamefont {J.}~\bibnamefont
  {Sous}}, \bibinfo {author} {\bibfnamefont {M.}~\bibnamefont {Chakraborty}},
  \bibinfo {author} {\bibfnamefont {R.~V.}\ \bibnamefont {Krems}}, \ and\
  \bibinfo {author} {\bibfnamefont {M.}~\bibnamefont {Berciu}},\ }\bibfield
  {title} {\enquote {\bibinfo {title} {Light bipolarons stabilized by {P}eierls
  electron-phonon coupling},}\ }\href
  {https://link.aps.org/doi/10.1103/PhysRevLett.121.247001} {\bibfield
  {journal} {\bibinfo  {journal} {Phys. Rev. Lett.}\ }\textbf {\bibinfo
  {volume} {121}},\ \bibinfo {pages} {247001} (\bibinfo {year}
  {2018})}\BibitemShut {NoStop}%
\bibitem [{\citenamefont {Zhang}\ \emph {et~al.}(2022)\citenamefont {Zhang},
  \citenamefont {Prokof’ev},\ and\ \citenamefont
  {Svistunov}}]{QMCBondBipolaron}%
  \BibitemOpen
  \bibfield  {author} {\bibinfo {author} {\bibfnamefont {C.}~\bibnamefont
  {Zhang}}, \bibinfo {author} {\bibfnamefont {N.~V.}\ \bibnamefont
  {Prokof’ev}}, \ and\ \bibinfo {author} {\bibfnamefont {B.~V.}\ \bibnamefont
  {Svistunov}},\ }\bibfield  {title} {\enquote {\bibinfo {title} {Bond
  bipolarons: {S}ign-free {M}onte {C}arlo approach},}\ }\href {\doibase
  10.1103/PhysRevB.105.L020501} {\bibfield  {journal} {\bibinfo  {journal}
  {Phys. Rev. B}\ }\textbf {\bibinfo {volume} {105}},\ \bibinfo {pages}
  {L020501} (\bibinfo {year} {2022})}\BibitemShut {NoStop}%
\bibitem [{\citenamefont {Fisher}\ and\ \citenamefont
  {Hohenberg}(1988)}]{LogBose2D1}%
  \BibitemOpen
  \bibfield  {author} {\bibinfo {author} {\bibfnamefont {D.~S.}\ \bibnamefont
  {Fisher}}\ and\ \bibinfo {author} {\bibfnamefont {P.~C.}\ \bibnamefont
  {Hohenberg}},\ }\bibfield  {title} {\enquote {\bibinfo {title} {Dilute {B}ose
  gas in two dimensions},}\ }\href {\doibase 10.1103/PhysRevB.37.4936}
  {\bibfield  {journal} {\bibinfo  {journal} {Phys. Rev. B}\ }\textbf {\bibinfo
  {volume} {37}},\ \bibinfo {pages} {4936} (\bibinfo {year}
  {1988})}\BibitemShut {NoStop}%
\bibitem [{\citenamefont {Prokof’ev}\ \emph {et~al.}(2001)\citenamefont
  {Prokof’ev}, \citenamefont {Ruebenacker},\ and\ \citenamefont
  {Svistunov}}]{LogBose2D2}%
  \BibitemOpen
  \bibfield  {author} {\bibinfo {author} {\bibfnamefont {N.}~\bibnamefont
  {Prokof’ev}}, \bibinfo {author} {\bibfnamefont {O.}~\bibnamefont
  {Ruebenacker}}, \ and\ \bibinfo {author} {\bibfnamefont {B.}~\bibnamefont
  {Svistunov}},\ }\bibfield  {title} {\enquote {\bibinfo {title} {Critical
  point of a weakly interacting two-dimensional {B}ose gas},}\ }\href {\doibase
  10.1103/PhysRevLett.87.270402} {\bibfield  {journal} {\bibinfo  {journal}
  {Phys. Rev. Lett.}\ }\textbf {\bibinfo {volume} {87}},\ \bibinfo {pages}
  {270402} (\bibinfo {year} {2001})}\BibitemShut {NoStop}%
\bibitem [{\citenamefont {Pilati}\ \emph {et~al.}(2008)\citenamefont {Pilati},
  \citenamefont {Giorgini},\ and\ \citenamefont {Prokof’ev}}]{LogBose2D3}%
  \BibitemOpen
  \bibfield  {author} {\bibinfo {author} {\bibfnamefont {S.}~\bibnamefont
  {Pilati}}, \bibinfo {author} {\bibfnamefont {S.}~\bibnamefont {Giorgini}}, \
  and\ \bibinfo {author} {\bibfnamefont {N.}~\bibnamefont {Prokof’ev}},\
  }\bibfield  {title} {\enquote {\bibinfo {title} {Critical temperature of
  interacting {B}ose gases in two and three dimensions},}\ }\href {\doibase
  10.1103/PhysRevLett.100.140405} {\bibfield  {journal} {\bibinfo  {journal}
  {Phys. Rev. Lett.}\ }\textbf {\bibinfo {volume} {100}},\ \bibinfo {pages}
  {140405} (\bibinfo {year} {2008})}\BibitemShut {NoStop}%
\bibitem [{\citenamefont {Nocera}\ \emph {et~al.}(2021)\citenamefont {Nocera},
  \citenamefont {Sous}, \citenamefont {Feiguin},\ and\ \citenamefont
  {Berciu}}]{NoceraPS}%
  \BibitemOpen
  \bibfield  {author} {\bibinfo {author} {\bibfnamefont {A.}~\bibnamefont
  {Nocera}}, \bibinfo {author} {\bibfnamefont {J.}~\bibnamefont {Sous}},
  \bibinfo {author} {\bibfnamefont {A.~E.}\ \bibnamefont {Feiguin}}, \ and\
  \bibinfo {author} {\bibfnamefont {M.}~\bibnamefont {Berciu}},\ }\bibfield
  {title} {\enquote {\bibinfo {title} {Bipolaron liquids at strong {P}eierls
  electron-phonon couplings},}\ }\href {\doibase 10.1103/PhysRevB.104.L201109}
  {\bibfield  {journal} {\bibinfo  {journal} {Phys. Rev. B}\ }\textbf {\bibinfo
  {volume} {104}},\ \bibinfo {pages} {L201109} (\bibinfo {year}
  {2021})}\BibitemShut {NoStop}%
\bibitem [{Note1()}]{Note1}%
  \BibitemOpen
  \bibinfo {note} {Since $T_\protect \mathrm {c}$ depends
  double-logarithmically weakly on the effective bipolaron-bipolaron
  interaction, neglecting deviations from the hard-core bipolaron-bipolaron
  interaction in our estimates of $T_\protect \mathrm {c}$ can only lead to
  small uncertainties~\cite {LogBose2D2,LogBose2D3}, which, nonetheless, do not
  affect any of our main results.}\BibitemShut {Stop}%
\bibitem [{\citenamefont {Macridin}(2003)}]{macridin2003phonons}%
  \BibitemOpen
  \bibfield  {author} {\bibinfo {author} {\bibfnamefont {A.}~\bibnamefont
  {Macridin}},\ }\href@noop {} {\emph {\bibinfo {title} {Ph. D. Thesis}}}\
  (\bibinfo  {publisher} {Rijksuniversiteit Groningen},\ \bibinfo {year}
  {2003})\BibitemShut {NoStop}%
\bibitem [{\citenamefont {McMillan}(1968)}]{McMillanFormula}%
  \BibitemOpen
  \bibfield  {author} {\bibinfo {author} {\bibfnamefont {W.~L.}\ \bibnamefont
  {McMillan}},\ }\bibfield  {title} {\enquote {\bibinfo {title} {Transition
  temperature of strong-coupled superconductors},}\ }\href {\doibase
  10.1103/PhysRev.167.331} {\bibfield  {journal} {\bibinfo  {journal} {Phys.
  Rev.}\ }\textbf {\bibinfo {volume} {167}},\ \bibinfo {pages} {331} (\bibinfo
  {year} {1968})}\BibitemShut {NoStop}%
\bibitem [{\citenamefont {Esterlis}\ \emph
  {et~al.}(2018{\natexlab{b}})\citenamefont {Esterlis}, \citenamefont
  {Kivelson},\ and\ \citenamefont {Scalapino}}]{TcBoundKivelson}%
  \BibitemOpen
  \bibfield  {author} {\bibinfo {author} {\bibfnamefont {I.}~\bibnamefont
  {Esterlis}}, \bibinfo {author} {\bibfnamefont {S.~A.}\ \bibnamefont
  {Kivelson}}, \ and\ \bibinfo {author} {\bibfnamefont {D.~J.}\ \bibnamefont
  {Scalapino}},\ }\bibfield  {title} {\enquote {\bibinfo {title} {A bound on
  the superconducting transition temperature},}\ }\href {\doibase
  10.1038/s41535-018-0133-0} {\bibfield  {journal} {\bibinfo  {journal} {npj
  Quantum Mater.}\ }\textbf {\bibinfo {volume} {3}},\ \bibinfo {pages} {1}
  (\bibinfo {year} {2018}{\natexlab{b}})}\BibitemShut {NoStop}%
\bibitem [{\citenamefont {Xing}\ \emph {et~al.}(2021)\citenamefont {Xing},
  \citenamefont {Chiu}, \citenamefont {Poletti}, \citenamefont {Scalettar},\
  and\ \citenamefont {Batrouni}}]{BoSSH2D}%
  \BibitemOpen
  \bibfield  {author} {\bibinfo {author} {\bibfnamefont {B.}~\bibnamefont
  {Xing}}, \bibinfo {author} {\bibfnamefont {W.-T.}\ \bibnamefont {Chiu}},
  \bibinfo {author} {\bibfnamefont {D.}~\bibnamefont {Poletti}}, \bibinfo
  {author} {\bibfnamefont {R.~T.}\ \bibnamefont {Scalettar}}, \ and\ \bibinfo
  {author} {\bibfnamefont {G.}~\bibnamefont {Batrouni}},\ }\bibfield  {title}
  {\enquote {\bibinfo {title} {Quantum {M}onte {C}arlo simulations of the 2{D}
  {S}u-{S}chrieffer-{H}eeger model},}\ }\href {\doibase
  10.1103/PhysRevLett.126.017601} {\bibfield  {journal} {\bibinfo  {journal}
  {Phys. Rev. Lett.}\ }\textbf {\bibinfo {volume} {126}},\ \bibinfo {pages}
  {017601} (\bibinfo {year} {2021})}\BibitemShut {NoStop}%
\bibitem [{\citenamefont {Cai}\ \emph {et~al.}(2021)\citenamefont {Cai},
  \citenamefont {Li},\ and\ \citenamefont {Yao}}]{CaiAFM}%
  \BibitemOpen
  \bibfield  {author} {\bibinfo {author} {\bibfnamefont {X.}~\bibnamefont
  {Cai}}, \bibinfo {author} {\bibfnamefont {Z.-X.}\ \bibnamefont {Li}}, \ and\
  \bibinfo {author} {\bibfnamefont {H.}~\bibnamefont {Yao}},\ }\bibfield
  {title} {\enquote {\bibinfo {title} {Antiferromagnetism induced by bond
  {S}u-{S}chrieffer-{H}eeger electron-phonon coupling: {A} quantum {M}onte
  {C}arlo study},}\ }\href {\doibase 10.1103/PhysRevLett.127.247203} {\bibfield
   {journal} {\bibinfo  {journal} {Phys. Rev. Lett.}\ }\textbf {\bibinfo
  {volume} {127}},\ \bibinfo {pages} {247203} (\bibinfo {year}
  {2021})}\BibitemShut {NoStop}%
\bibitem [{\citenamefont {G\"otz}\ \emph {et~al.}(2022)\citenamefont {G\"otz},
  \citenamefont {Beyl}, \citenamefont {Hohenadler},\ and\ \citenamefont
  {Assaad}}]{Assaad2D}%
  \BibitemOpen
  \bibfield  {author} {\bibinfo {author} {\bibfnamefont {A.}~\bibnamefont
  {G\"otz}}, \bibinfo {author} {\bibfnamefont {S.}~\bibnamefont {Beyl}},
  \bibinfo {author} {\bibfnamefont {M.}~\bibnamefont {Hohenadler}}, \ and\
  \bibinfo {author} {\bibfnamefont {F.~F.}\ \bibnamefont {Assaad}},\ }\bibfield
   {title} {\enquote {\bibinfo {title} {Valence-bond solid to antiferromagnet
  transition in the two-dimensional {S}u-{S}chrieffer-{H}eeger model by
  {L}angevin dynamics},}\ }\href {\doibase 10.1103/PhysRevB.105.085151}
  {\bibfield  {journal} {\bibinfo  {journal} {Phys. Rev. B}\ }\textbf {\bibinfo
  {volume} {105}},\ \bibinfo {pages} {085151} (\bibinfo {year}
  {2022})}\BibitemShut {NoStop}%
\bibitem [{\citenamefont {Feng}\ \emph {et~al.}(2021)\citenamefont {Feng},
  \citenamefont {Xing}, \citenamefont {Poletti}, \citenamefont {Scalettar},\
  and\ \citenamefont {Batrouni}}]{ScaletterSSHU}%
  \BibitemOpen
  \bibfield  {author} {\bibinfo {author} {\bibfnamefont {C.}~\bibnamefont
  {Feng}}, \bibinfo {author} {\bibfnamefont {B.}~\bibnamefont {Xing}}, \bibinfo
  {author} {\bibfnamefont {D.}~\bibnamefont {Poletti}}, \bibinfo {author}
  {\bibfnamefont {R.}~\bibnamefont {Scalettar}}, \ and\ \bibinfo {author}
  {\bibfnamefont {G.}~\bibnamefont {Batrouni}},\ }\bibfield  {title} {\enquote
  {\bibinfo {title} {Phase diagram of the {S}u-{S}chrieffer-{H}eeger-{H}ubbard
  model on a square lattice},}\ }\href@noop {} {\bibfield  {journal} {\bibinfo
  {journal} {arXiv:2109.09206}\ } (\bibinfo {year} {2021})}\BibitemShut
  {NoStop}%
\bibitem [{\citenamefont {Yin}\ \emph {et~al.}(2011)\citenamefont {Yin},
  \citenamefont {Haule},\ and\ \citenamefont {Kotliar}}]{HauleNatMater}%
  \BibitemOpen
  \bibfield  {author} {\bibinfo {author} {\bibfnamefont {Z.~P.}\ \bibnamefont
  {Yin}}, \bibinfo {author} {\bibfnamefont {K.}~\bibnamefont {Haule}}, \ and\
  \bibinfo {author} {\bibfnamefont {G.}~\bibnamefont {Kotliar}},\ }\bibfield
  {title} {\enquote {\bibinfo {title} {Kinetic frustration and the nature of
  the magnetic and paramagnetic states in iron pnictides and iron
  chalcogenides},}\ }\href {\doibase 10.1038/nmat3120} {\bibfield  {journal}
  {\bibinfo  {journal} {Nat. Mater.}\ }\textbf {\bibinfo {volume} {10}},\
  \bibinfo {pages} {932} (\bibinfo {year} {2011})}\BibitemShut {NoStop}%
\bibitem [{\citenamefont {Zhang}\ \emph {et~al.}(2014)\citenamefont {Zhang},
  \citenamefont {Harriger}, \citenamefont {Yin}, \citenamefont {Lv},
  \citenamefont {Wang}, \citenamefont {Tan}, \citenamefont {Song},
  \citenamefont {Abernathy}, \citenamefont {Tian}, \citenamefont {Egami},
  \citenamefont {Haule}, \citenamefont {Kotliar},\ and\ \citenamefont
  {Dai}}]{PnictogenHeight}%
  \BibitemOpen
  \bibfield  {author} {\bibinfo {author} {\bibfnamefont {C.}~\bibnamefont
  {Zhang}}, \bibinfo {author} {\bibfnamefont {L.~W.}\ \bibnamefont {Harriger}},
  \bibinfo {author} {\bibfnamefont {Z.}~\bibnamefont {Yin}}, \bibinfo {author}
  {\bibfnamefont {W.}~\bibnamefont {Lv}}, \bibinfo {author} {\bibfnamefont
  {M.}~\bibnamefont {Wang}}, \bibinfo {author} {\bibfnamefont {G.}~\bibnamefont
  {Tan}}, \bibinfo {author} {\bibfnamefont {Y.}~\bibnamefont {Song}}, \bibinfo
  {author} {\bibfnamefont {D.~L.}\ \bibnamefont {Abernathy}}, \bibinfo {author}
  {\bibfnamefont {W.}~\bibnamefont {Tian}}, \bibinfo {author} {\bibfnamefont
  {T.}~\bibnamefont {Egami}}, \bibinfo {author} {\bibfnamefont
  {K.}~\bibnamefont {Haule}}, \bibinfo {author} {\bibfnamefont
  {G.}~\bibnamefont {Kotliar}}, \ and\ \bibinfo {author} {\bibfnamefont
  {P.}~\bibnamefont {Dai}},\ }\bibfield  {title} {\enquote {\bibinfo {title}
  {Effect of pnictogen height on spin waves in iron pnictides},}\ }\href
  {\doibase 10.1103/PhysRevLett.112.217202} {\bibfield  {journal} {\bibinfo
  {journal} {Phys. Rev. Lett.}\ }\textbf {\bibinfo {volume} {112}},\ \bibinfo
  {pages} {217202} (\bibinfo {year} {2014})}\BibitemShut {NoStop}%
\bibitem [{\citenamefont {Clay}\ and\ \citenamefont
  {Roy}(2020)}]{ModelQuarterFilling}%
  \BibitemOpen
  \bibfield  {author} {\bibinfo {author} {\bibfnamefont {R.~T.}\ \bibnamefont
  {Clay}}\ and\ \bibinfo {author} {\bibfnamefont {D.}~\bibnamefont {Roy}},\
  }\bibfield  {title} {\enquote {\bibinfo {title} {Superconductivity due to
  cooperation of electron-electron and electron-phonon interactions at quarter
  filling},}\ }\href {\doibase 10.1103/PhysRevResearch.2.023006} {\bibfield
  {journal} {\bibinfo  {journal} {Phys. Rev. Research}\ }\textbf {\bibinfo
  {volume} {2}},\ \bibinfo {pages} {023006} (\bibinfo {year}
  {2020})}\BibitemShut {NoStop}%
\bibitem [{\citenamefont {Lau}\ \emph {et~al.}(2007)\citenamefont {Lau},
  \citenamefont {Berciu},\ and\ \citenamefont {Sawatzky}}]{BreathingMode}%
  \BibitemOpen
  \bibfield  {author} {\bibinfo {author} {\bibfnamefont {B.}~\bibnamefont
  {Lau}}, \bibinfo {author} {\bibfnamefont {M.}~\bibnamefont {Berciu}}, \ and\
  \bibinfo {author} {\bibfnamefont {G.~A.}\ \bibnamefont {Sawatzky}},\
  }\bibfield  {title} {\enquote {\bibinfo {title} {Single-polaron properties of
  the one-dimensional breathing-mode {H}amiltonian},}\ }\href {\doibase
  10.1103/PhysRevB.76.174305} {\bibfield  {journal} {\bibinfo  {journal} {Phys.
  Rev. B}\ }\textbf {\bibinfo {volume} {76}},\ \bibinfo {pages} {174305}
  (\bibinfo {year} {2007})}\BibitemShut {NoStop}%
\bibitem [{\citenamefont {Werman}\ \emph {et~al.}(2017)\citenamefont {Werman},
  \citenamefont {Kivelson},\ and\ \citenamefont {Berg}}]{KivelsonPhononLargeN}%
  \BibitemOpen
  \bibfield  {author} {\bibinfo {author} {\bibfnamefont {Y.}~\bibnamefont
  {Werman}}, \bibinfo {author} {\bibfnamefont {S.~A.}\ \bibnamefont
  {Kivelson}}, \ and\ \bibinfo {author} {\bibfnamefont {E.}~\bibnamefont
  {Berg}},\ }\bibfield  {title} {\enquote {\bibinfo {title} {Non-quasiparticle
  transport and resistivity saturation: a view from the large-${N}$ limit},}\
  }\href {\doibase 10.1038/s41535-017-0009-8} {\bibfield  {journal} {\bibinfo
  {journal} {npj Quantum Mater.}\ }\textbf {\bibinfo {volume} {2}},\ \bibinfo
  {pages} {7} (\bibinfo {year} {2017})}\BibitemShut {NoStop}%
\bibitem [{\citenamefont {Rice}\ \emph {et~al.}(1993)\citenamefont {Rice},
  \citenamefont {Gopalan},\ and\ \citenamefont {Sigrist}}]{90bond}%
  \BibitemOpen
  \bibfield  {author} {\bibinfo {author} {\bibfnamefont {T.~M.}\ \bibnamefont
  {Rice}}, \bibinfo {author} {\bibfnamefont {S.}~\bibnamefont {Gopalan}}, \
  and\ \bibinfo {author} {\bibfnamefont {M.}~\bibnamefont {Sigrist}},\
  }\bibfield  {title} {\enquote {\bibinfo {title} {Superconductivity, spin gaps
  and {L}uttinger liquids in a class of cuprates},}\ }\href {\doibase
  10.1209/0295-5075/23/6/011} {\bibfield  {journal} {\bibinfo  {journal}
  {Europhys. Lett.}\ }\textbf {\bibinfo {volume} {23}},\ \bibinfo {pages} {445}
  (\bibinfo {year} {1993})}\BibitemShut {NoStop}%
\bibitem [{\citenamefont {Shen}\ \emph {et~al.}(2002)\citenamefont {Shen},
  \citenamefont {Lanzara}, \citenamefont {Ishihara},\ and\ \citenamefont
  {Nagaosa}}]{ZXNagaosaPeierlsCuprates}%
  \BibitemOpen
  \bibfield  {author} {\bibinfo {author} {\bibfnamefont {Z.-X.}\ \bibnamefont
  {Shen}}, \bibinfo {author} {\bibfnamefont {A.}~\bibnamefont {Lanzara}},
  \bibinfo {author} {\bibfnamefont {S.}~\bibnamefont {Ishihara}}, \ and\
  \bibinfo {author} {\bibfnamefont {N.}~\bibnamefont {Nagaosa}},\ }\bibfield
  {title} {\enquote {\bibinfo {title} {Role of the electron-phonon interaction
  in the strongly correlated cuprate superconductors},}\ }\href {\doibase
  10.1080/13642810208220725} {\bibfield  {journal} {\bibinfo  {journal}
  {Philos. Mag. B}\ }\textbf {\bibinfo {volume} {82}},\ \bibinfo {pages} {1349}
  (\bibinfo {year} {2002})}\BibitemShut {NoStop}%
\bibitem [{\citenamefont {Prodi}\ \emph {et~al.}(2004)\citenamefont {Prodi},
  \citenamefont {Gilioli}, \citenamefont {Gauzzi}, \citenamefont {Licci},
  \citenamefont {Marezio}, \citenamefont {Bolzoni}, \citenamefont {Huang},
  \citenamefont {Santoro},\ and\ \citenamefont {Lynn}}]{corner-sharing}%
  \BibitemOpen
  \bibfield  {author} {\bibinfo {author} {\bibfnamefont {A.}~\bibnamefont
  {Prodi}}, \bibinfo {author} {\bibfnamefont {E.}~\bibnamefont {Gilioli}},
  \bibinfo {author} {\bibfnamefont {A.}~\bibnamefont {Gauzzi}}, \bibinfo
  {author} {\bibfnamefont {F.}~\bibnamefont {Licci}}, \bibinfo {author}
  {\bibfnamefont {M.}~\bibnamefont {Marezio}}, \bibinfo {author} {\bibfnamefont
  {F.}~\bibnamefont {Bolzoni}}, \bibinfo {author} {\bibfnamefont
  {Q.}~\bibnamefont {Huang}}, \bibinfo {author} {\bibfnamefont
  {A.}~\bibnamefont {Santoro}}, \ and\ \bibinfo {author} {\bibfnamefont
  {J.~W.}\ \bibnamefont {Lynn}},\ }\bibfield  {title} {\enquote {\bibinfo
  {title} {Charge, orbital and spin ordering phenomena in the mixed valence
  manganite
  ({N}a{M}n$^{3+}$$_{3}$)({M}n$^{3+}$$_{2}${M}n$^{4+}$$_{2}$){O}$_{12}$},}\
  }\href {\doibase 10.1038/nmat1038} {\bibfield  {journal} {\bibinfo  {journal}
  {Nat. Mater.}\ }\textbf {\bibinfo {volume} {3}},\ \bibinfo {pages} {48}
  (\bibinfo {year} {2004})}\BibitemShut {NoStop}%
\bibitem [{\citenamefont {Gerber}\ \emph {et~al.}(2017)\citenamefont {Gerber},
  \citenamefont {Yang}, \citenamefont {Zhu}, \citenamefont {Soifer},
  \citenamefont {Sobota}, \citenamefont {Rebec}, \citenamefont {Lee},
  \citenamefont {Jia}, \citenamefont {Moritz}, \citenamefont {Jia},
  \citenamefont {Gauthier}, \citenamefont {Li}, \citenamefont {Leuenberger},
  \citenamefont {Zhang}, \citenamefont {Chaix}, \citenamefont {Li},
  \citenamefont {Jang}, \citenamefont {Lee}, \citenamefont {Yi}, \citenamefont
  {Dakovski}, \citenamefont {Song}, \citenamefont {Glownia}, \citenamefont
  {Nelson}, \citenamefont {Kim}, \citenamefont {Chuang}, \citenamefont
  {Hussain}, \citenamefont {Moore}, \citenamefont {Devereaux}, \citenamefont
  {Lee}, \citenamefont {Kirchmann},\ and\ \citenamefont {Shen}}]{ZXSHENFESE}%
  \BibitemOpen
  \bibfield  {author} {\bibinfo {author} {\bibfnamefont {S.}~\bibnamefont
  {Gerber}}, \bibinfo {author} {\bibfnamefont {S.-L.}\ \bibnamefont {Yang}},
  \bibinfo {author} {\bibfnamefont {D.}~\bibnamefont {Zhu}}, \bibinfo {author}
  {\bibfnamefont {H.}~\bibnamefont {Soifer}}, \bibinfo {author} {\bibfnamefont
  {J.~A.}\ \bibnamefont {Sobota}}, \bibinfo {author} {\bibfnamefont
  {S.}~\bibnamefont {Rebec}}, \bibinfo {author} {\bibfnamefont {J.~J.}\
  \bibnamefont {Lee}}, \bibinfo {author} {\bibfnamefont {T.}~\bibnamefont
  {Jia}}, \bibinfo {author} {\bibfnamefont {B.}~\bibnamefont {Moritz}},
  \bibinfo {author} {\bibfnamefont {C.}~\bibnamefont {Jia}}, \bibinfo {author}
  {\bibfnamefont {A.}~\bibnamefont {Gauthier}}, \bibinfo {author}
  {\bibfnamefont {Y.}~\bibnamefont {Li}}, \bibinfo {author} {\bibfnamefont
  {D.}~\bibnamefont {Leuenberger}}, \bibinfo {author} {\bibfnamefont
  {Y.}~\bibnamefont {Zhang}}, \bibinfo {author} {\bibfnamefont
  {L.}~\bibnamefont {Chaix}}, \bibinfo {author} {\bibfnamefont
  {W.}~\bibnamefont {Li}}, \bibinfo {author} {\bibfnamefont {H.}~\bibnamefont
  {Jang}}, \bibinfo {author} {\bibfnamefont {J.-S.}\ \bibnamefont {Lee}},
  \bibinfo {author} {\bibfnamefont {M.}~\bibnamefont {Yi}}, \bibinfo {author}
  {\bibfnamefont {G.~L.}\ \bibnamefont {Dakovski}}, \bibinfo {author}
  {\bibfnamefont {S.}~\bibnamefont {Song}}, \bibinfo {author} {\bibfnamefont
  {J.~M.}\ \bibnamefont {Glownia}}, \bibinfo {author} {\bibfnamefont
  {S.}~\bibnamefont {Nelson}}, \bibinfo {author} {\bibfnamefont {K.~W.}\
  \bibnamefont {Kim}}, \bibinfo {author} {\bibfnamefont {Y.-D.}\ \bibnamefont
  {Chuang}}, \bibinfo {author} {\bibfnamefont {Z.}~\bibnamefont {Hussain}},
  \bibinfo {author} {\bibfnamefont {R.~G.}\ \bibnamefont {Moore}}, \bibinfo
  {author} {\bibfnamefont {T.~P.}\ \bibnamefont {Devereaux}}, \bibinfo {author}
  {\bibfnamefont {W.-S.}\ \bibnamefont {Lee}}, \bibinfo {author} {\bibfnamefont
  {P.~S.}\ \bibnamefont {Kirchmann}}, \ and\ \bibinfo {author} {\bibfnamefont
  {Z.-X.}\ \bibnamefont {Shen}},\ }\bibfield  {title} {\enquote {\bibinfo
  {title} {Femtosecond electron-phonon lock-in by photoemission and x-ray
  free-electron laser},}\ }\href {\doibase 10.1126/science.aak9946} {\bibfield
  {journal} {\bibinfo  {journal} {Science}\ }\textbf {\bibinfo {volume}
  {357}},\ \bibinfo {pages} {71} (\bibinfo {year} {2017})}\BibitemShut
  {NoStop}%
\bibitem [{\citenamefont {Ahn}\ \emph {et~al.}(2021)\citenamefont {Ahn},
  \citenamefont {Cavalleri}, \citenamefont {Georges}, \citenamefont
  {Ismail-Beigi}, \citenamefont {Millis},\ and\ \citenamefont
  {Triscone}}]{FunctionalTMOxides}%
  \BibitemOpen
  \bibfield  {author} {\bibinfo {author} {\bibfnamefont {C.}~\bibnamefont
  {Ahn}}, \bibinfo {author} {\bibfnamefont {A.}~\bibnamefont {Cavalleri}},
  \bibinfo {author} {\bibfnamefont {A.}~\bibnamefont {Georges}}, \bibinfo
  {author} {\bibfnamefont {S.}~\bibnamefont {Ismail-Beigi}}, \bibinfo {author}
  {\bibfnamefont {A.~J.}\ \bibnamefont {Millis}}, \ and\ \bibinfo {author}
  {\bibfnamefont {J.-M.}\ \bibnamefont {Triscone}},\ }\bibfield  {title}
  {\enquote {\bibinfo {title} {Designing and controlling the properties of
  transition metal oxide quantum materials},}\ }\href {\doibase
  10.1038/s41563-021-00989-2} {\bibfield  {journal} {\bibinfo  {journal} {Nat.
  Mater.}\ }\textbf {\bibinfo {volume} {1}} (\bibinfo {year} {2021}),\
  10.1038/s41563-021-00989-2}\BibitemShut {NoStop}%
\bibitem [{\citenamefont {Kennes}\ \emph {et~al.}(2021)\citenamefont {Kennes},
  \citenamefont {Claassen}, \citenamefont {Xian}, \citenamefont {Georges},
  \citenamefont {Millis}, \citenamefont {Hone}, \citenamefont {Dean},
  \citenamefont {Basov}, \citenamefont {Pasupathy},\ and\ \citenamefont
  {Rubio}}]{MoireFunctional}%
  \BibitemOpen
  \bibfield  {author} {\bibinfo {author} {\bibfnamefont {D.~M}\ \bibnamefont
  {Kennes}}, \bibinfo {author} {\bibfnamefont {M.}~\bibnamefont {Claassen}},
  \bibinfo {author} {\bibfnamefont {L.}~\bibnamefont {Xian}}, \bibinfo {author}
  {\bibfnamefont {A.}~\bibnamefont {Georges}}, \bibinfo {author} {\bibfnamefont
  {A.~J.}\ \bibnamefont {Millis}}, \bibinfo {author} {\bibfnamefont
  {J.}~\bibnamefont {Hone}}, \bibinfo {author} {\bibfnamefont {C.~R.}\
  \bibnamefont {Dean}}, \bibinfo {author} {\bibfnamefont {D.~N.}\ \bibnamefont
  {Basov}}, \bibinfo {author} {\bibfnamefont {A.~N}\ \bibnamefont {Pasupathy}},
  \ and\ \bibinfo {author} {\bibfnamefont {A.}~\bibnamefont {Rubio}},\
  }\bibfield  {title} {\enquote {\bibinfo {title} {Moir{\'e} heterostructures
  as a condensed-matter quantum simulator},}\ }\href {\doibase
  https://doi.org/10.1038/s41567-020-01154-3} {\bibfield  {journal} {\bibinfo
  {journal} {Nat. Phys.}\ }\textbf {\bibinfo {volume} {17}},\ \bibinfo {pages}
  {155} (\bibinfo {year} {2021})}\BibitemShut {NoStop}%
\bibitem [{\citenamefont {Cao}\ \emph {et~al.}(2018)\citenamefont {Cao},
  \citenamefont {Fatemi}, \citenamefont {Fang}, \citenamefont {Watanabe},
  \citenamefont {Taniguchi}, \citenamefont {Kaxiras},\ and\ \citenamefont
  {Jarillo-Herrero}}]{MAG}%
  \BibitemOpen
  \bibfield  {author} {\bibinfo {author} {\bibfnamefont {Y.}~\bibnamefont
  {Cao}}, \bibinfo {author} {\bibfnamefont {V.}~\bibnamefont {Fatemi}},
  \bibinfo {author} {\bibfnamefont {S.}~\bibnamefont {Fang}}, \bibinfo {author}
  {\bibfnamefont {K.}~\bibnamefont {Watanabe}}, \bibinfo {author}
  {\bibfnamefont {T.}~\bibnamefont {Taniguchi}}, \bibinfo {author}
  {\bibfnamefont {E.}~\bibnamefont {Kaxiras}}, \ and\ \bibinfo {author}
  {\bibfnamefont {P.}~\bibnamefont {Jarillo-Herrero}},\ }\bibfield  {title}
  {\enquote {\bibinfo {title} {Unconventional superconductivity in magic-angle
  graphene superlattices},}\ }\href {\doibase 10.1038/nature26160} {\bibfield
  {journal} {\bibinfo  {journal} {Nature}\ }\textbf {\bibinfo {volume} {556}},\
  \bibinfo {pages} {43} (\bibinfo {year} {2018})}\BibitemShut {NoStop}%
\bibitem [{\citenamefont {Doud}\ \emph {et~al.}(2020)\citenamefont {Doud},
  \citenamefont {Voevodin}, \citenamefont {Hochuli}, \citenamefont
  {Champsaur},\ and\ \citenamefont {Nuckolls}}]{SuperatomicMaterials}%
  \BibitemOpen
  \bibfield  {author} {\bibinfo {author} {\bibfnamefont {E.~A.}\ \bibnamefont
  {Doud}}, \bibinfo {author} {\bibfnamefont {A.}~\bibnamefont {Voevodin}},
  \bibinfo {author} {\bibfnamefont {T.~J.}\ \bibnamefont {Hochuli}}, \bibinfo
  {author} {\bibfnamefont {A.~M.}\ \bibnamefont {Champsaur}}, \ and\ \bibinfo
  {author} {\bibfnamefont {C.}~\bibnamefont {Nuckolls}},\ }\bibfield  {title}
  {\enquote {\bibinfo {title} {Superatoms in materials science},}\ }\href
  {\doibase 10.1038/s41578-019-0175-3} {\bibfield  {journal} {\bibinfo
  {journal} {Nat. Rev. Mater.}\ }\textbf {\bibinfo {volume} {5}} (\bibinfo
  {year} {2020}),\ 10.1038/s41578-019-0175-3}\BibitemShut {NoStop}%
\bibitem [{\citenamefont {Prokof’ev}\ \emph
  {et~al.}(1998{\natexlab{a}})\citenamefont {Prokof’ev}, \citenamefont
  {Svistunov},\ and\ \citenamefont {Tupitsyn}}]{Worm1}%
  \BibitemOpen
  \bibfield  {author} {\bibinfo {author} {\bibfnamefont {N.~V.}\ \bibnamefont
  {Prokof’ev}}, \bibinfo {author} {\bibfnamefont {B.~V.}\ \bibnamefont
  {Svistunov}}, \ and\ \bibinfo {author} {\bibfnamefont {I.~S.}\ \bibnamefont
  {Tupitsyn}},\ }\bibfield  {title} {\enquote {\bibinfo {title} {Exact,
  complete, and universal continuous-time worldline {M}onte {C}arlo approach to
  the statistics of discrete quantum systems},}\ }\href@noop {} {\bibfield
  {journal} {\bibinfo  {journal} {Sov. Phys. JETP}\ }\textbf {\bibinfo {volume}
  {87}},\ \bibinfo {pages} {310} (\bibinfo {year}
  {1998}{\natexlab{a}})}\BibitemShut {NoStop}%
\bibitem [{\citenamefont {Prokof’ev}\ \emph
  {et~al.}(1998{\natexlab{b}})\citenamefont {Prokof’ev}, \citenamefont
  {Svistunov},\ and\ \citenamefont {Tupitsyn}}]{Worm2}%
  \BibitemOpen
  \bibfield  {author} {\bibinfo {author} {\bibfnamefont {N.~V.}\ \bibnamefont
  {Prokof’ev}}, \bibinfo {author} {\bibfnamefont {B.~V.}\ \bibnamefont
  {Svistunov}}, \ and\ \bibinfo {author} {\bibfnamefont {I.~S.}\ \bibnamefont
  {Tupitsyn}},\ }\bibfield  {title} {\enquote {\bibinfo {title} {“{W}orm”
  algorithm in quantum {M}onte {C}arlo simulations},}\ }\href {\doibase
  https://doi.org/10.1016/S0375-9601(97)00957-2} {\bibfield  {journal}
  {\bibinfo  {journal} {Phys. Lett. A}\ }\textbf {\bibinfo {volume} {238}},\
  \bibinfo {pages} {253} (\bibinfo {year} {1998}{\natexlab{b}})}\BibitemShut
  {NoStop}%
\bibitem [{\citenamefont {Boninsegni}\ and\ \citenamefont
  {Ceperley}(1995)}]{massQMCuniversal}%
  \BibitemOpen
  \bibfield  {author} {\bibinfo {author} {\bibfnamefont {M.}~\bibnamefont
  {Boninsegni}}\ and\ \bibinfo {author} {\bibfnamefont {D.~M.}\ \bibnamefont
  {Ceperley}},\ }\bibfield  {title} {\enquote {\bibinfo {title} {Path integral
  {M}onte {C}arlo simulation of isotopic liquid helium mixtures},}\ }\href
  {\doibase 10.1103/PhysRevLett.74.2288} {\bibfield  {journal} {\bibinfo
  {journal} {Phys. Rev. Lett.}\ }\textbf {\bibinfo {volume} {74}},\ \bibinfo
  {pages} {2288} (\bibinfo {year} {1995})}\BibitemShut {NoStop}%
\bibitem [{\citenamefont {Bergmann}\ and\ \citenamefont
  {Rainer}(1973)}]{Bergmann73}%
  \BibitemOpen
  \bibfield  {author} {\bibinfo {author} {\bibfnamefont {G.}~\bibnamefont
  {Bergmann}}\ and\ \bibinfo {author} {\bibfnamefont {D.}~\bibnamefont
  {Rainer}},\ }\bibfield  {title} {\enquote {\bibinfo {title} {The sensitivity
  of the transition temperature to changes in $\alpha^2f(\omega)$},}\ }\href
  {\doibase https://doi.org/10.1007/BF02351862} {\bibfield  {journal} {\bibinfo
   {journal} {Z. Phys.}\ }\textbf {\bibinfo {volume} {263}},\ \bibinfo {pages}
  {59} (\bibinfo {year} {1973})}\BibitemShut {NoStop}%
\bibitem [{\citenamefont {Nosarzewski}\ \emph {et~al.}(2021)\citenamefont
  {Nosarzewski}, \citenamefont {Huang}, \citenamefont {Dee}, \citenamefont
  {Esterlis}, \citenamefont {Moritz}, \citenamefont {Kivelson}, \citenamefont
  {Johnston},\ and\ \citenamefont {Devereaux}}]{GeneralHolsteinBipolaron}%
  \BibitemOpen
  \bibfield  {author} {\bibinfo {author} {\bibfnamefont {B.}~\bibnamefont
  {Nosarzewski}}, \bibinfo {author} {\bibfnamefont {E.~W.}\ \bibnamefont
  {Huang}}, \bibinfo {author} {\bibfnamefont {Philip~M.}\ \bibnamefont {Dee}},
  \bibinfo {author} {\bibfnamefont {I.}~\bibnamefont {Esterlis}}, \bibinfo
  {author} {\bibfnamefont {B.}~\bibnamefont {Moritz}}, \bibinfo {author}
  {\bibfnamefont {S.~A.}\ \bibnamefont {Kivelson}}, \bibinfo {author}
  {\bibfnamefont {S.}~\bibnamefont {Johnston}}, \ and\ \bibinfo {author}
  {\bibfnamefont {T.~P.}\ \bibnamefont {Devereaux}},\ }\bibfield  {title}
  {\enquote {\bibinfo {title} {Superconductivity, charge density waves, and
  bipolarons in the {H}olstein model},}\ }\href {\doibase
  10.1103/PhysRevB.103.235156} {\bibfield  {journal} {\bibinfo  {journal}
  {Phys. Rev. B}\ }\textbf {\bibinfo {volume} {103}},\ \bibinfo {pages}
  {235156} (\bibinfo {year} {2021})}\BibitemShut {NoStop}%
\bibitem [{\citenamefont {Takahashi}(1977)}]{Takahashi_1977}%
  \BibitemOpen
  \bibfield  {author} {\bibinfo {author} {\bibfnamefont {M}~\bibnamefont
  {Takahashi}},\ }\bibfield  {title} {\enquote {\bibinfo {title} {Half-filled
  {H}ubbard model at low temperature},}\ }\href {\doibase
  10.1088/0022-3719/10/8/031} {\bibfield  {journal} {\bibinfo  {journal} {J.
  Phys.}\ }\textbf {\bibinfo {volume} {10}},\ \bibinfo {pages} {1289} (\bibinfo
  {year} {1977})}\BibitemShut {NoStop}%
\bibitem [{\citenamefont {Sous}\ \emph {et~al.}(2017)\citenamefont {Sous},
  \citenamefont {Chakraborty}, \citenamefont {Adolphs}, \citenamefont {Krems},\
  and\ \citenamefont {Berciu}}]{SousScRep}%
  \BibitemOpen
  \bibfield  {author} {\bibinfo {author} {\bibfnamefont {J.}~\bibnamefont
  {Sous}}, \bibinfo {author} {\bibfnamefont {M.}~\bibnamefont {Chakraborty}},
  \bibinfo {author} {\bibfnamefont {C.~P.~J.}\ \bibnamefont {Adolphs}},
  \bibinfo {author} {\bibfnamefont {R.~V.}\ \bibnamefont {Krems}}, \ and\
  \bibinfo {author} {\bibfnamefont {M.}~\bibnamefont {Berciu}},\ }\bibfield
  {title} {\enquote {\bibinfo {title} {Phonon-mediated repulsion, sharp
  transitions and (quasi)self-trapping in the extended {P}eierls-{H}ubbard
  model},}\ }\href {\doibase 10.1038/s41598-017-01228-y} {\bibfield  {journal}
  {\bibinfo  {journal} {Sci. Rep.}\ }\textbf {\bibinfo {volume} {7}},\ \bibinfo
  {pages} {1} (\bibinfo {year} {2017})}\BibitemShut {NoStop}%
\bibitem [{\citenamefont {Kagan}\ and\ \citenamefont {Klinger}(1976)}]{Kagan}%
  \BibitemOpen
  \bibfield  {author} {\bibinfo {author} {\bibfnamefont {Y.}~\bibnamefont
  {Kagan}}\ and\ \bibinfo {author} {\bibfnamefont {M.~I.}\ \bibnamefont
  {Klinger}},\ }\bibfield  {title} {\enquote {\bibinfo {title} {Role of
  fluctuational barrier “preparation” in the quantum diffusion of atomic
  particles in a crystal},}\ }\href@noop {} {\bibfield  {journal} {\bibinfo
  {journal} {Zh. Eksp. Teor. Fiz.}\ }\textbf {\bibinfo {volume} {70}},\
  \bibinfo {pages} {255} (\bibinfo {year} {1976})}\BibitemShut {NoStop}%
\bibitem [{\citenamefont {Lee}\ \emph {et~al.}(2014)\citenamefont {Lee},
  \citenamefont {Schmitt}, \citenamefont {Moore}, \citenamefont {Johnston},
  \citenamefont {Cui}, \citenamefont {Li}, \citenamefont {Yi}, \citenamefont
  {Liu}, \citenamefont {Hashimoto}, \citenamefont {Zhang}, \citenamefont {Lu},
  \citenamefont {Devereaux}, \citenamefont {Lee},\ and\ \citenamefont
  {Shen}}]{ZXSHENFESE0}%
  \BibitemOpen
  \bibfield  {author} {\bibinfo {author} {\bibfnamefont {J.~J.}\ \bibnamefont
  {Lee}}, \bibinfo {author} {\bibfnamefont {F.~T.}\ \bibnamefont {Schmitt}},
  \bibinfo {author} {\bibfnamefont {R.~G.}\ \bibnamefont {Moore}}, \bibinfo
  {author} {\bibfnamefont {S.}~\bibnamefont {Johnston}}, \bibinfo {author}
  {\bibfnamefont {Y.-T.}\ \bibnamefont {Cui}}, \bibinfo {author} {\bibfnamefont
  {W.}~\bibnamefont {Li}}, \bibinfo {author} {\bibfnamefont {M.}~\bibnamefont
  {Yi}}, \bibinfo {author} {\bibfnamefont {Z.~K.}\ \bibnamefont {Liu}},
  \bibinfo {author} {\bibfnamefont {M.}~\bibnamefont {Hashimoto}}, \bibinfo
  {author} {\bibfnamefont {Y.}~\bibnamefont {Zhang}}, \bibinfo {author}
  {\bibfnamefont {D.~H.}\ \bibnamefont {Lu}}, \bibinfo {author} {\bibfnamefont
  {T.~P.}\ \bibnamefont {Devereaux}}, \bibinfo {author} {\bibfnamefont {D.-H.}\
  \bibnamefont {Lee}}, \ and\ \bibinfo {author} {\bibfnamefont {Z.-X.}\
  \bibnamefont {Shen}},\ }\bibfield  {title} {\enquote {\bibinfo {title}
  {Interfacial mode coupling as the origin of the enhancement of
  ${T}_\mathrm{c}$ in {F}e{S}e films on {S}r{T}i{O}$_3$},}\ }\href {\doibase
  10.1038/nature13894} {\bibfield  {journal} {\bibinfo  {journal} {Nature}\
  }\textbf {\bibinfo {volume} {515}},\ \bibinfo {pages} {245} (\bibinfo {year}
  {2014})}\BibitemShut {NoStop}%
\end{thebibliography}

\providecommand{\noopsort}[1]{}\providecommand{\singleletter}[1]{#1}%

\end{document}